\documentclass[aip,pop,reprint,numerical,twocolumn]{revtex4-1}

\usepackage{commands}

\usepackage{circuitikz}
\usepackage{tabularx}

\begin{document}

\title{Exploring the parameter space of MagLIF implosions using similarity scaling.~~I.~Theoretical framework}
\date{\today}
\author{D.~E.~Ruiz}
\email{deruiz@sandia.gov}
\affiliation{Sandia National Laboratories, P.O. Box 5800, Albuquerque, NM 87185, USA}
\author{P.~F.~Schmit}
\affiliation{Lawrence Livermore National Laboratory, Livermore, CA 94550, USA}
\author{D.~A.~Yager-Elorriaga}
\affiliation{Sandia National Laboratories, P.O. Box 5800, Albuquerque, NM 87185, USA}
\author{C.~A.~Jennings}
\affiliation{Sandia National Laboratories, P.O. Box 5800, Albuquerque, NM 87185, USA}
\author{K.~Beckwith}
\affiliation{Sandia National Laboratories, P.O. Box 5800, Albuquerque, NM 87185, USA}

%%%%%%%%%%%%%%%%%%%%%%%%%%%%%%%%%%%%%%%%%%%%%%%%%
%%%%%%%%%%%%%%%%%%%%%%%%%%%%%%%%%%%%%%%%%%%%%%%%%
%%%%%%%%%%%%%%%%%%%%%%%%%%%%%%%%%%%%%%%%%%%%%%%%%

\begin{abstract}

Magneto-inertial fusion (MIF) concepts, such as the Magnetized Liner Inertial Fusion (MagLIF) platform [M.~ R.~Gomez \etal, Phys. Rev. Lett. \textbf{113}, 155003 (2014)], constitute a promising path for achieving ignition and significant fusion yields in the laboratory.  The space of experimental input parameters defining a MagLIF load is highly multi-dimensional, and the implosion itself is a complex event involving many physical processes.   In the first paper of this series, we develop a simplified analytical model that identifies the main physical processes at play during a MagLIF implosion.  Using non-dimensional analysis, we determine the most important dimensionless parameters characterizing MagLIF implosions and provide estimates of such parameters using typical fielded or experimentally observed quantities for MagLIF.  We then show that MagLIF loads can be ``incompletely" similarity scaled, meaning that the experimental input parameters of MagLIF can be varied such that many (but not all) of the dimensionless quantities are conserved.  Based on similarity-scaling arguments, we can explore the parameter space of MagLIF loads and estimate the performance of the scaled loads.  In the follow-up papers of this series, we test the similar scaling theory for MagLIF loads against simulations for two different scaling ``vectors", which include current scaling and rise-time scaling.

\end{abstract}

%\pacs{52.35.Fp, 47.10.ab, 52.25.-b, 52.65.Rr}

\maketitle

%%%%%%%%%%%%%%%%%%%%%%%%%%%%%%%%%%%%%%%%%%%%%%%%%
%%%%%%%%%%%%%%%%%%%%%%%%%%%%%%%%%%%%%%%%%%%%%%%%%
%%%%%%%%%%%%%%%%%%%%%%%%%%%%%%%%%%%%%%%%%%%%%%%%%
\section{Introduction}

For the last 60 years, one of the main approaches to produce thermonuclear self-heating plasmas has been laser-driven inertial confinement fusion (ICF).\cite{Lindl:2004jz,Hurricane:2019hd}  In this approach, powerful lasers directly or indirectly energize the outside surface of spherical capsules to achieve high ablation pressures ($\sim100$~Mbar) and implosion velocities greater than $\sim400$ km/s.  Recent experiments on the National Ignition Facility (NIF) have demonstrated hot-spot ignition using indirect drive.\cite{PhysRevLett.129.075001,PhysRevE.106.025202,PhysRevE.106.025201}  Given the limited energy that the NIF can deliver to a fuel capsule, hot-spot ignition attempts to minimize the required laser energy by igniting a central hot spot in the fuel, which then propagates into the dense outer fuel layer.  However, hot-spot ignition places high engineering requirements on the manufacturing of the fuel capsule, on the design and delivery of the laser-pulse shape, and on the symmetry of the laser drive to maximize compression of the hot spot.  As an alternative pathway to achieving ignition and high fusion yields in the laboratory, magneto-inertial fusion (MIF) relaxes these stringent implosion conditions by introducing strong magnetic fields, which modify the transport properties of the fuel and permit slower implosions.\cite{Widner_1977,Lindemuth_1981,Lindemuth:1983aa,Lindemuth:2015fu}  In particular, pulsed-power driven MIF is a technological approach that utilizes energy-rich, pulsed-power generators to compress a load configuration to thermonuclear conditions, albeit at a longer implosion timescale compared to laser-driven ICF.

One interesting MIF concept is the Magnetized Liner Inertial Fusion (MagLIF) platform,\cite{Slutz:2010hd,Gomez:2014eta} which is being studied at the Z Pulsed-Power Facility of Sandia National Laboratories.\cite{Sefkow:2014ik,Gomez:2019bg,Knapp:2019gf,Sinars:2020bv,YagerElorriaga:2022cp}  The Z facility delivers a $\sim$20-MA current pulse to the cylindrical MagLIF load, which then implodes under the action of the $\vec{J}\times \vec{B}$ force.\cite{Gomez:2020cd}  Since MagLIF employs a relatively thick and heavy metallic cylindrical tamper, or liner, the implosion velocities are $70-100$ km/s, which are substantially lower than those in traditional ICF.  {\color{black}In MagLIF, shock heating of the fuel is negligible.  Therefore, to raise the initial fuel adiabat prior to compression, the fuel is preheated by a 2-4 kJ, TW-class laser.\cite{Weis:2021id,HarveyThompson:2020jr,HarveyThompson:2019ff,HarveyThompson:2018dd,Geissel:2018eea,HarveyThompson:2016iw,HarveyThompson:2015jv}}  Since the implosions are slower (on the order of 100 ns), the fuel must be premagnetized to reduce thermal conduction losses.\cite{Slutz:2010hd}  This is achieved by external electromagnetic coils that premagnetize the entire fuel volume with an axial 10--16 T magnetic field.\cite{Gomez:2020cd}  The combination of these elements has led to thermonuclear yield production in laboratory experiments\cite{Gomez:2014eta,Gomez:2019bg,Gomez:2020cd} and significant plasma magnetization inferred via secondary DT neutron emission.\cite{Schmit:2014fg,Knapp:2015kc,Lewis:2021kz}

Pulsed-power driven MagLIF loads are complex systems that involve multiple processes that govern the electrical-circuit dynamics, liner-implosion dynamics, hydrodynamic instabilities, fuel-energy losses, particle transport, fuel-mass losses, axial magnetic-field transport, fusion yield, etc.  Also, the experimental input parameters defining a MagLIF load are numerous.  Main input parameters include: (i)~the liner geometry (inner radius, outer radius, and height), (ii)~the material composition of the liner, (iii)~the preheat energy, (iv)~the initial fuel density, (v)~the external axial magnetic field, (vi)~the peak current delivered to the load, and (vii)~the rise time of the current waveform delivered to the load.  Overall, the input-parameter space defining a MagLIF load is relatively high dimensional, and a detailed exploration of such space is therefore difficult.

Testing MagLIF performance in the entire input-parameter space is virtually impossible via integrated experiments on Z.  Radiation-hydrodynamic simulations offer an alternative to survey the performance of MagLIF, but such a study would still be computationally expensive when considering a large portion of the input-parameter space.  Moreover, given the complexity of MagLIF implosions, there is also a level of uncertainty when surveying areas in parameter space that have not yet been experimentally fielded, even with today's highly complex multi-physics codes.\cite{Slutz:2010hd,Sefkow:2014ik,HarveyThompson:2018dd,Weis:2021id}  The exploration of the parameter space for MagLIF can be simplified {\color{black}via \textit{dimensional analysis}} and \textit{similarity scaling}.

Our goal in this series of papers is to introduce a theoretical framework for similarity scaling MagLIF loads.  This framework provides analytical scaling prescriptions for the main input parameters defining a MagLIF load and also gives estimates of performance of the resulting implosions.  The theory also provides guidance on what simulations should be done and what experiments should be fielded on Z.

In Paper I of this series, we develop a simplified analytical model based on ordinary differential equations (ODEs) that contains the main physical processes at play during a MagLIF implosion.  In many aspects, this model can be considered as a further reduced version of the ODE-based model presented in \Refs{McBride:2015ga,McBride:2016dm}.  After introducing the governing equations and writing them in dimensionless form, we determine the most important dimensionless parameters characterizing MagLIF implosions.  We also provide estimates of such parameters using typical fielded or experimentally observed stagnation quantities for MagLIF.  As originally  demonstrated in \Refa{Schmit:2020jd}, we then show that MagLIF loads can be \textit{incompletely similarly scaled}, meaning that the experimental input parameters of MagLIF can be varied such that \textit{many but not all} of the dimensionless quantities characterizing a MagLIF implosion are conserved.  Following arguments based on similarity scaling, we can then estimate the performance of scaled MagLIF loads when using this framework. 

Specific applications of these results are left to \Refs{foot:Ruiz_current,foot:Ruiz_time} (also referred as Papers II and III).  In  Paper II, we address the problem of scaling MagLIF loads with respect to the peak electrical current.  In Paper III, we study the scaling MagLIF loads when the characteristic time of the voltage drive (or the current-rise time) is varied.

This paper can be divided into two (informal) parts.  {\color{black}The first part comprises Secs.~\ref{sec:electrical}--\ref{sec:summary},} which discuss the main physical processes at play in a MagLIF implosion.  In these sections, we identify the dimensionless parameters characterizing such processes.  {\color{black}In \Sec{sec:electrical}, we introduce a circuit model to describe the coupling between the pulsed-power generator and the z-pinch MagLIF load.}  In \Sec{sec:liner}, we present a simplified two-layer liner model.  In \Sec{sec:stability}, we introduce a model for the liner in-flight aspect ratio, and we show how it can serve as a measure of the robustness of MagLIF liners towards the magneto-Rayleigh--Taylor (MRT) instability.  In \Sec{sec:energy}, we introduce a model describing the fuel energy balance.  In \Sec{sec:mass}, we discuss fuel mass losses as the liner implodes.  In \Sec{sec:Bz}, we discuss the main processes at play during the flux compression of the axial magnetic field inside the fuel.  In \Sec{sec:neutron}, we introduce the basic model for the fusion yield.  In \Sec{sec:summary}, we summarize the dimensionless parameters characterizing the identified main physical processes occurring in a MagLIF implosion and evaluate them using typical fielded or experimentally observed  quantities for MagLIF.

The second part of this paper comprises \Sec{sec:scaling}.  Here we discuss similarity scaling applied to MagLIF, and we introduce the notions of scale invariants, essential and non-essential dimensionless variables, and incomplete similarity scaling.  We present the three major assumptions underlying the similarity-scaling framework applied to MagLIF and derive the general similarity-scaling prescriptions for the experimental input parameters characterizing a MagLIF configuration.  A reader who is more interested in the application of similarity scaling to MagLIF may begin by reading \Sec{sec:scaling} and then refer to the previous Secs.~\ref{sec:electrical}--\ref{sec:neutron} for more details.

%%%%%%%%%%%%%%%%%%%%%%%%%%%%%%%%%%%%%%%%%%%%%%%%%
\section{Electrical circuit model}
\label{sec:electrical}

In contrast to laser-driven ICF, where the laser output is independent of the imploding spherical shell, magnetically-driven fusion loads, such as MagLIF, are inherently coupled to the pulsed-power driver delivering the current.  Due to this dynamical coupling, it is necessary to model the dynamics of both the pulsed-power driver and the imploding load.

\begin{figure}
	\begin{circuitikz}[american,scale=1] \draw
	(0,3) to[sinusoidal voltage source,l=$\varphi_{\rm oc}$,i=$I_s$] (0,0)
	(0,3) to[resistor,l=$Z_0$] (2,3)
  			to[inductor,l=$L_0$] (3.5,3) --(5,3)
  		    to[inductor,l=$L_1$] (8,3)
  		    to[vL,l_=$L_{\rm load}$,i=$I_l$] (8,0) -- (0,0) 
    (3.5,3) to[capacitor,l^=$C$,v=$\varphi_c$] (3.5,0)
    (5,0) to[vR,l_=$R_{\rm loss}$] (5,3);
	\end{circuitikz}
	\caption{Representative circuit diagram of the Z generator.  Typical parameters used are $Z_0=0.18~\Omega$, $L_0 \simeq 8.34$~nH, $C \simeq 8.41$~nF, and $L_1 \simeq 5$~nH.}
	\label{fig:electrical:circuit}
\end{figure}
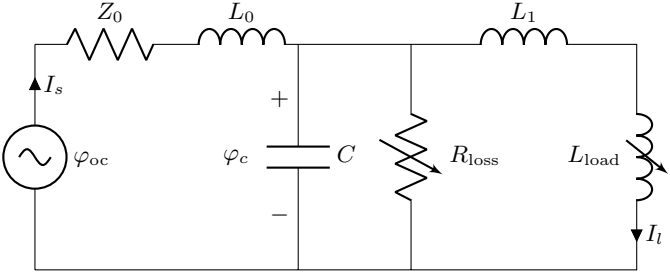

\begin{figure}
	\includegraphics[scale=.43]{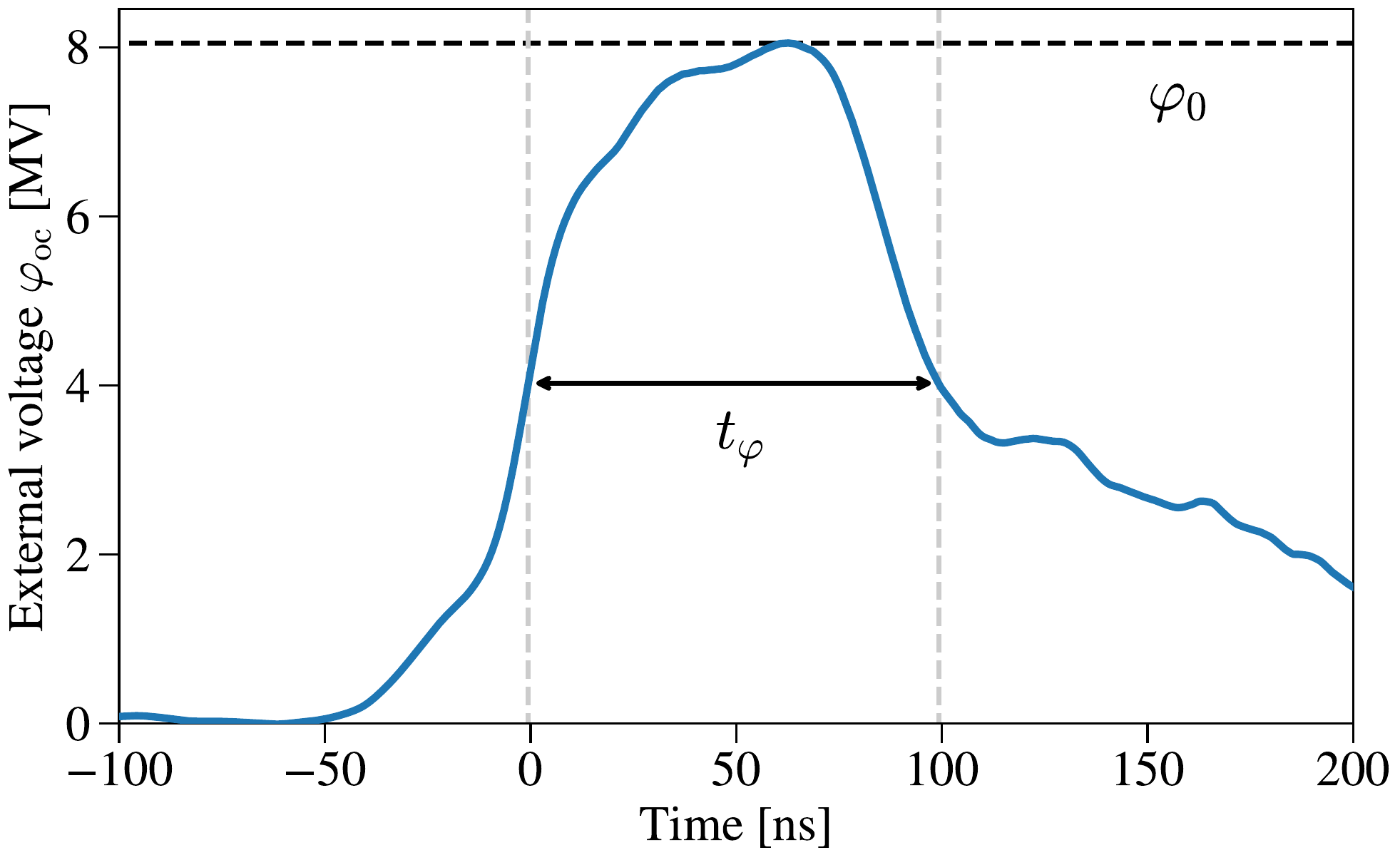}
	\caption{Example of a voltage source $\varphi_{\rm oc}$ as a function of time.  The characteristic voltage $\varphi_0$ can be taken as the maximum value of $\varphi_{\rm oc}$.  The characteristic time $t_\varphi$ is defined as the full-width half-maximum (FWHM) of the voltage curve.}
	\label{fig:electrical:Voc}
\end{figure}

The Z Pulsed-Power Facility can be described using the equivalent circuit model shown in \Fig{fig:electrical:circuit}.\cite{McBride:2010hb,McBride:2015ga}  The circuit is externally driven by a time-varying voltage $\varphi_{\rm oc}(t)$, which is twice the forward-going voltage at the vacuum-insulator stack on Z, is independent of the load, and usually does not vary much between experiments for a given charge voltage.  An example waveform for $\varphi_{\rm oc}(t)$ is shown in \Fig{fig:electrical:Voc}.

For the left-most circuit loop in \Fig{fig:electrical:circuit}, the governing equation for the source current $I_s(t)$ is
\begin{equation}
	L_0 \frac{\mathrm{d}}{\mathrm{d}t} I_s 
		= \varphi_{\rm oc} - Z_0 I_s - \varphi_c. 
	\label{eq:electrical:Is}
\end{equation}
Here $Z_0$ is the effective impedance of the Z accelerator.  $L_0$ and $C$ respectively denote the inductance and capacitance of the magnetically-insulated transmission lines (MITLs).  Finally, $\varphi_c(t)$ is the corresponding voltage across the capacitor $C$ associated to the MITLs.

For the circuit loop located at the center of \Fig{fig:electrical:circuit}, the governing equation is
\begin{equation}
	C \frac{\mathrm{d}}{\mathrm{d}t} \varphi_c 
		= I_s - I_l - \frac{\varphi_c}{R_{\rm loss}},
	\label{eq:electrical:Vc}
\end{equation}
where $I_l(t)$ is the electrical current going into the load region and $R_{\rm loss}(t)$ is a shunt resistor used to model losses at the post-hole convolute.\cite{Gomez:2017dw}  Here we model the shunt resistor using a prescribed time-dependent model so that $R_{\rm loss}$ is given by
\begin{equation}
	R_{\rm loss} (t)  \doteq R_{\rm loss,i} + (R_{\rm loss,f}- R_{\rm loss,i}) f(t),
	\label{eq:electric:Rloss}
\end{equation}
where
\begin{equation}
	f(t) = \frac{1}{1+ \exp\left( - \frac{t-t_{\rm loss}}{\Delta t_{\rm loss}} \right) } .
	\label{eq:electrical:f}
\end{equation}
is a function describing the transition from the initial loss resistance $R_{\rm loss,i}$ of the circuit early in time to the final loss resistance $R_{\rm loss,f}$ at later times.  Typically, $R_{\rm loss,i} > R_{\rm loss,f}$ meaning that the fraction of current losses to the load increases as time progresses.   In \Eq{eq:electrical:f}, $t_{\rm loss}$ represents the time at which current losses turn on, and $\Delta t_{\rm loss}$ is the characteristic time describing the transition in resistance of $R_{\rm loss}$. 

In \Fig{fig:electrical:circuit}, the load current $I_l$ is determined by
\begin{equation}
	\frac{\mathrm{d}}{\mathrm{d}t} \left( L_{\rm load} I_l \right) 
		+ L_1 \frac{\mathrm{d}}{\mathrm{d}t} I_l
		= \varphi_c,
	\label{eq:electrical:Il}
\end{equation}
where $L_1$ is the initial, time-independent inductance of the feed region, which includes the post-hole convolute, the inner-MITL, and the return-can volume.  $L_{\rm load}(t)$ is the time-dependent inductance of the MagLIF liner.  We approximate $L_{\rm load}$ as a coaxial inductance so that
\begin{equation}
	L_{\rm load} (t) \doteq  \frac{\mu_0 h}{2 \pi} \ln \left( \frac{R_{\rm out,0}}{R_{\rm out}} \right),
	\label{eq:electrical:Lload}
\end{equation}
where $R_{\rm out}(t)$ is the liner outer radius, $R_{\rm out,0}\doteq R_{\rm out}(0)$ is the initial outer radius, $h$ is the height of the imploding region of the liner, and $\mu_0$ is the magnetic permeability in vacuum.  For the sake of simplicity, we consider the liner as perfectly conducting so the driving azimuthal magnetic field does not diffuse into the liner.  Magnetic diffusion leads to Ohmic heating of the liner surface and adds a (small) additional inductance term associated to the magnetic flux contained in the liner region.  The former acts as an energy source for the liner material and can modify its effective compressibility.  A more detailed model that includes magnetic diffusion is provided in \Refa{McBride:2015ga}. 

Let us now write the circuit equations \eq{eq:electrical:Is}--\eq{eq:electrical:Il} in dimensionless form and identify the dimensionless groups that characterize the system.  We introduce the following dimensionless variables:
\begin{equation}
	\begin{aligned}
	\bar{t}_i &=\frac{t_i}{t_{\rm \varphi}},  & \bar{I}_s & = \frac{I_s}{I_\star},  &
	\bar{I}_l & = \frac{I_l}{I_\star} ,  \\
   \bar{\varphi}_{\rm oc} &= \frac{\varphi_{\rm oc}}{\varphi_0}, &
	\qquad \bar{\varphi}_c & = \frac{\varphi_c}{\varphi_0},	 &
	\bar{R}_{\rm out} & = \frac{R_{\rm out}}{R_{\rm out,0}}.
	\end{aligned}
	\label{eq:electrical:variables}
\end{equation}
(From hereon, the ``barred" quantities $\bar{Q}$ denote dimensionless quantities.)  Here $\varphi_0$ is the  characteristic amplitude of the voltage source, which we define as the maximum voltage.  The parameter $t_{\rm \varphi}$ represents the characteristic timescale of the voltage source.  For pulsed-power systems, the maximum voltage and the duration of the voltage source near peak voltage are two key factors determining the current delivered to the load.  We thus define $t_{\rm \varphi}$ as the full-width half-maximum (FWHM) of the voltage curve;  \eg, see \Fig{fig:electrical:Voc}.  In terms of the normalized voltage trace $\bar{\varphi}_{\rm oc}(\bar{t})$, we can then write the external voltage source as $\varphi_{\rm oc}(t) = \varphi_0 \bar{\varphi}_{\rm oc}(t/t_{\rm \varphi})$.  In \Eq{eq:electrical:variables},  $I_\star$ is the characteristic current of the circuit.  To find a suitable expression for $I_\star$, we consider a short-circuit configuration in \Fig{fig:electrical:circuit} with no dynamic load; \ie $L_{\rm load} =0$.  In the absence of the MITL capacitance $C$ and current losses, the characteristic voltages across the voltage source $\varphi_{\rm oc}$, the impedance $Z_0$, and the inductors $L_0$ and $L_1$ satisfy
\begin{equation}
	\varphi_0 \sim Z_0 I_\star + \frac{L_0 + L_1}{t_\varphi} I_\star.
\end{equation}
Therefore, we define $I_\star \doteq \varphi_0 [Z_0 +(L_0 +L_1)/t_\varphi]^{-1}$, which in convenient units is written as
\begin{equation}
	I_\star \mathrm{(MA)}
		\doteq  \frac{[\varphi_0({\rm MV})] }
					{ [Z_0({\rm Ohm}) ]  
					  + \frac{[L_0(\mathrm{nH})] + [L_1(\mathrm{nH})]}{ [t_{\rm \varphi}({\rm ns}) ] } 
					 }.
	\label{eq:electrical:Istar}
\end{equation}
Finally, $t_i$ in \Eq{eq:electrical:variables} denotes the explicit time parameters appearing in \Eqs{eq:electrical:Is}--\eq{eq:electrical:Il}, such as $t_{\rm loss}$ and $\Delta t_{\rm loss}$.

We substitute \Eqs{eq:electrical:variables} into \Eqs{eq:electrical:Is}--\eq{eq:electrical:Il}.  The resulting dimensionless equations are the following:
\begin{gather}
	\frac{c_1}{1+c_2} \frac{\mathrm{d}}{\mathrm{d} \bar{t}} \, \bar{I}_s 
		 = - \frac{c_2}{1+c_2}	\, \bar{I}_s + \bar{\varphi}_{\rm oc} - \bar{\varphi}_c , 
	\label{eq:electrical:Is_bar} \\
	c_3^2 \frac{\mathrm{d}}{\mathrm{d} \bar{t}} \, \bar{\varphi}_c 
		 = \frac{\bar{I}_s - \bar{I}_l}{1+c_2} - \frac{\bar{\varphi}_c}{c_4 + (c_5-c_4)f},
	\label{eq:electrical:Vc_bar} \\
	\left(1-c_1 - c_6 \ln \bar{R}_{\rm out} \right) \frac{\mathrm{d}}{\mathrm{d} \bar{t}} \, \bar{I}_l 
		 = c_6 \, \frac{\bar{I}_l}{\bar{R}_{\rm out}} \frac{\mathrm{d} \bar{R}_{\rm out}}{\mathrm{d} \bar{t}}   + (1+c_2) \bar{\varphi}_c.
	\label{eq:electrical:Il_bar}
\end{gather}
In terms of convenient units, the dimensionless parameters appearing in \Eqs{eq:electrical:Is_bar}--\eq{eq:electrical:Il_bar} may be written as
\begin{align}
	c_1 & \doteq \frac{[L_0(\mathrm{nH})]} {[L_0(\mathrm{nH})] + [L_1(\mathrm{nH})] }, 
		\label{eq:electrical:c1} \\
	c_2 & \doteq \frac{[Z_0(\mathrm{Ohm})] \cdot 
							[t_{\rm \varphi}~(\mathrm{ns})]} {[L_0(\mathrm{nH})] + [L_1(\mathrm{nH})]} ,
		\label{eq:electrical:c2} \\
	c_3 & \doteq   \sqrt{ [L_0(\mathrm{nH})]+[L_1(\mathrm{nH})] }  \cdot \frac{  [C (\mathrm{nF})]^{1/2} } 
							{ [t_{\rm \varphi}~(\mathrm{ns})] },
		\label{eq:electrical:c3} \\
	c_4 & \doteq \frac{[R_{\rm loss,i}(\mathrm{Ohm})] \cdot 
							[t_{\rm \varphi}~(\mathrm{ns})] } {[L_0(\mathrm{nH})] + [L_1(\mathrm{nH})] }, 
		\label{eq:electrical:c4} \\
	c_5 & \doteq \frac{[R_{\rm loss,f}(\mathrm{Ohm})] \cdot 
							[t_{\rm \varphi}~(\mathrm{ns})] } { [L_0(\mathrm{nH})] + [L_1(\mathrm{nH})] }, 
		\label{eq:electrical:c5} \\
	c_6 & \doteq  \frac{\mu_0 h}{2\pi (L_0+L_1)} = 2 \frac{[h(\mathrm{cm})]} { [L_0(\mathrm{nH})] + [L_1(\mathrm{nH})] }.
		\label{eq:electrical:c6} 
\end{align}

The dimensionless parameters $c_1$ to $c_6$ characterize the circuit dynamics.  In \Eq{eq:electrical:c1}, the parameter $c_1$ is the ratio of the inductance $L_0$ of the outer MITLs and the total initial inductance of the ``shorted circuit" $(L_0+L_1)$.  This parameter measures the static-inductance matching between the pulsed-power driver and the initial feed inductance (convolute, inner MITL and load).  The parameter $c_2$ in \Eq{eq:electrical:c2} is the ratio of the timescale $t_\varphi$ of the voltage drive and the characteristic time $(L_0+L_1)/Z_0$ of the pulsed-power generator.  This parameter is a measure of the resonance between the voltage drive and the pulsed-power generator and serves as a good indicator of the driving efficiency.  The parameter $c_3$ in \Eq{eq:electrical:c3} is the ratio of the time $\sqrt{(L_0+L_1) C}$ and the characteristic time $t_\varphi$.  This parameter measures the coupling of the circuit to the capacitance $C$ of the MITLs; in other words, it describes how well the MITLs can capacitively store and release electrical energy into the circuit.  The parameters $c_4$ and $c_5$ measure how efficient the current transmission is into the load region early and late in time.  When $c_4$ or $c_5$ are large, there are negligible current losses; when $c_4$ or $c_5$ are small, current losses can be significant.  {\color{black}Finally, the dimensionless parameter $c_6$ in \Eq{eq:electrical:c6} and the parameter $\Pi$ defined afterwards in \Eq{eq:liner:Pi} characterize the coupling between the time-dependent inductance of the imploding load and the rest of the circuit [see \Eq{eq:electrical:Il_bar}].  When $c_6$ and $\Pi$ are large, the coupling is significant meaning that changes in the time-dependent load  inductance affect the delivered current to the load.  As shown in \Eq{eq:electrical:c6}, the parameter $c_6$ increases when the imploding height $h$ of the liner is larger and when the initial total inductance $(L_0+L_1)$ of the circuit is smaller, which is the case of low-impedance pulsed-power generators.}

%%%%%%%%%%%%%%%%%%%%%%%%%%%%%%%%%%%%%%%%%%%%%%%%%
\section{Liner dynamics}
\label{sec:liner}

To describe the implosion of the liner, we introduce the following simple \textit{two-layer} liner model.  The governing equations for the inner and outer radii of the liner are
\begin{gather}
	\frac{\widehat{m}}{2}	\, \frac{\mathrm{d}^2}{\mathrm{d} t^2} R_{\rm out}
			= 2\pi R_{\rm out}  \left( p_{\rm liner} - p_{\rm mag,ext} \right), 
	\label{eq:liner:Rout}	\\
	\frac{\widehat{m}}{2}	\, \frac{\mathrm{d}^2}{\mathrm{d} t^2} R_{\rm in}
			= 2\pi R_{\rm in}  \left( p_{\rm fuel}	 + p_{\rm mag,int} - p_{\rm liner}\right). 
	\label{eq:liner:Rin}
\end{gather}  
Here $R_{\rm out}(t) $ and $R_{\rm in}(t)$ are the outer and inner radii of the liner, respectively.  To each shell, we assign half of the liner mass per-unit-length $\widehat{m}$,\cite{foot:mass_partition} which is defined as
\begin{equation}
	\widehat{m} = \pi \rho_{\rm liner,0} \left (R_{\rm out,0}^2 - R_{\rm in,0}^2\right),
	\label{eq:liner:m}
\end{equation}
where $\rho_{\rm liner,0}$ is the initial mass density of the liner, $R_{\rm out,0}$ is the initial liner outer radius, and $R_{\rm in,0}$ is the initial liner inner radius.  It is customary to use the liner aspect ratio 
\begin{equation}
	{\rm AR} \doteq \frac{R_{\rm out,0}}{\delta R_0 } = \frac{R_{\rm out,0}}{R_{\rm out,0}-R_{\rm in,0}}
\end{equation}
where $\delta R_0$ is the initial liner thickness.  For large initial AR liners, \ie thin shells, the liner mass per-unit-length is approximately $\widehat{m}\simeq 2\pi \rho_{\rm liner,0} R_{\rm out,0}^2/ \mathrm{AR}$.

The liner is subject to the internal and external magnetic-field pressures, the internal pressure within the liner, and the pressure of the fuel inside the liner cavity.  The external magnetic pressure $p_{\rm mag,ext}(t)$ is given by
\begin{equation}
	p_{\rm mag,ext}	\doteq \frac{B_\theta^2}{2\mu_0} = \frac{\mu_0 I_l^2}{8\pi^2 R_{\rm out}^2},
 	\label{eq:liner:pmag}
\end{equation}
where $B_\theta (t)$ is the azimuthal magnetic field outside the liner generated by the load current $I_l$.

We assume that the internal pressure $p_{\rm liner}(t)$ of the liner is solely determined by the mass density of the liner at a given time.  In other words, we assume that the liner pressure obeys a cold adiabatic equation of state so that
\begin{equation}
	p_{\rm liner} 
		\doteq		
			p_{\rm ref} \left( \frac{\rho_{\rm liner}}{\rho_{\rm ref}} \right)^\gamma
		= 	p_{\rm ref} \left( \frac{\widehat{m}/\rho_{\rm ref} }
				{ \pi (R_{\rm out}^2 - R_{\rm in}^2 )} \right)^\gamma.
	\label{eq:liner:pliner}
\end{equation}
Here $p_{\rm ref} $, $\rho_{\rm ref}$, and $\gamma$ are the reference pressure, mass density, and polytropic index that characterize the adiabatically compressed liner material.  We note that, during a MagLIF implosion, the liner accesses diverse regions of the material-EOS phase space.  The trajectories in the EOS phase space cannot be entirely described by the simple adiabatic-compression model \eq{eq:liner:pliner} since a MagLIF liner is usually shocked by the driving magnetic pressure early in the implosion.  Moreover, the EOS phase-space trajectories depend on the liner-implosion dynamics and can change when scaling MagLIF loads.  Although \Eq{eq:liner:pliner} has its limitations, we choose this model for the liner EOS in order to simplify the upcoming dimensional analysis.  Revisions could be incorporated to include a more complete physics picture.  For example, the predictive capability of the model could be increased by obtaining the parameters $p_{\rm ref} $, $\rho_{\rm ref}$, and $\gamma$ from the regions in phase space that a MagLIF implosion accesses in more sophisticated material EOS models.  Improvements such as this one will be left for future work.
  
%We use this model for the liner EOS in order to simplify the upcoming dimensional analysis.   Although this model has its limitations, revisions could be incorporated to include a more complete physics picture.  For example, the predictive capability of this model could be increased by obtaining the parameters $p_{\rm ref} $, $\rho_{\rm ref}$, and $\gamma$ from the regions in phase space that a MagLIF implosion accesses in more sophisticated material EOS models.  Improvements such as this one will be left for future work.

%In this work, we consider Be liners.  Upon fitting a power law to a cold equation of state\cite{McBride:2015ga} for Be within the 1000-3000 Mbar range (typical stagnation pressures), we obtain  $p_{\rm ref} \doteq 66.91$~Mbar, $\rho_{\rm ref}\doteq 14$~g/cc, and $\gamma\doteq 1.923$.  These values will be used throughout this work.   

In \Eq{eq:liner:Rin}, $p_{\rm mag,int}(t)$ denotes the magnetic pressure due to the flux-compressed axial magnetic field $B_z(t)$ within the fuel.  It is a given by
\begin{equation}
	p_{\rm mag,int}	\doteq \frac{B_z^2}{2\mu_0}.
 	\label{eq:liner:pmagz}
\end{equation}
A model for $B_z(t)$ will be introduced in \Sec{sec:Bz}.

In \Eq{eq:liner:Rin}, $p_{\rm fuel}(t)$ denotes the internal pressure of the compressed fuel.  {\color{black}For hydrogen-like plasma with charge number $Z=1$,} the fuel pressure is
\begin{equation}
	p_{\rm fuel}	
		=	2 \frac{ \rho }{m_i} k_B T
		=	\frac{2}{3} \frac{U}{\pi R_{\rm in}^2 h},
	\label{eq:liner:pfuel}
\end{equation}
where $\rho(t)$ is the fuel mass density, $T(t)$ is the fuel temperature,\cite{foot:temp} and $U(t)$ is the fuel internal energy.  Also, $k_B$ is the Boltzmann constant, $m_i\doteq Am_p$ is the ion mass, $A$ is the average atomic number for the fusion fuel ($A=$ for DD plasma and $A=2.5$ for equimolar DT plasma), and $m_p$ is the proton mass.  Dynamical equations for $U$ and $\rho$ are derived in Secs.~\ref{sec:energy} and \ref{sec:mass}, respectively.

As a side note, it is common to describe imploding z-pinch liners as thin shells.\cite{Ryutov:2000zz,Schmit:2020jd}  The thin-shell model can be readily retrieved by summing \Eqs{eq:liner:Rout} and \eq{eq:liner:Rin} and approximating $R(t) \simeq R_{\rm out}(t) \simeq R_{\rm in}(t)$.  This leads to
\begin{equation}
	\widehat{m}	\frac{\mathrm{d}^2}{\mathrm{d} t^2} R
			\simeq 2\pi R  \left( p_{\rm fuel}	 + p_{\rm mag,int} - p_{\rm mag,ext} \right).
	\label{eq:basic:thin_shell}
\end{equation}
This model is appropriate when the shell thickness $\delta R(t) \doteq R_{\rm out}(t) - R_{\rm in}(t)$ is much smaller than the outer or inner shell radii.  For  high-AR liners, it satisfactorily represents the initial dynamics of the implosion.  However, it is no longer a good approximation near stagnation, when the liner thickness is almost equivalent to the liner outer radius; \ie when $R_{\rm in}\simeq 0$ and $R_{\rm out}\simeq \delta R$.

We now determine the main dimensionless parameters characterizing the liner implosion.  We introduce the following additional dimensionless variables:
\begin{equation}
	\begin{aligned}
 	\bar{R}_{\rm in} & \doteq \frac{R_{\rm in}}{R_{\rm in,0}},&
 	\bar{p}_{\rm fuel} & \doteq \frac{p_{\rm fuel}	}{p_{\rm preheat} }, &
	\bar{B}_z & \doteq \frac{B_z}{B_{z,0}} ,
	\end{aligned}
	\label{eq:liner:variables2}
\end{equation}
where $B_{z,0}$ is the initial externally-applied axial magnetic field and $p_{\rm preheat}$ is the characteristic fuel pressure that is achieved after the end of the preheat stage.  In terms of the initial liner dimensions and the total amount of energy $E_{\rm preheat}$ delivered to the fuel, $p_{\rm preheat}$ is given by
\begin{equation}
	p_{\rm preheat} 
		\doteq \frac{2}{3} \frac{E_{\rm preheat}}{\pi R_{\rm in ,0}^2 h}.
	\label{eq:liner:p_preheat}
\end{equation}
%
%In terms of convenient variables, we write the preheat pressure as
%
%\begin{equation}
%	p_{\rm preheat} {\rm (kbar)}
%		\doteq 2.12  \, 
%				\frac{[E_{\rm preheat} (\rm{kJ})]}{[R_{\rm in,0} ({\rm cm})]^2 \cdot [h ({\rm cm})]}.
%\end{equation}
%
%I think it’s better to present p_preheat as the characteristic pressure scale you get when you treat the liner initial dimensions and the total preheat energy as fundamental input quantities.  The way it’s phrased in the manuscript, it sounds like there are approximations being made that imply the theoretical model is only correct if preheat is introduced before the liner implodes significantly.  Your model is comprehensive, it’s only the interpretation of p_preheat that has an approximate translation.
%
In dimensionless form, \Eqs{eq:liner:Rout} and \eq{eq:liner:Rin} become
\begin{widetext}
\begin{align}
	\frac{1}{2} \frac{\mathrm{d}^2 \bar{R}_{\rm out}}{\mathrm{d} \bar{t}^2}
			= 	& 	- 	\Pi \frac{\bar{I}_l^2}{\bar{R}_{\rm out}} 
					+ 2 \Lambda \bar{R}_{\rm out}
					\left( \frac{\widehat{m}/\rho_{\rm ref}}{\pi R_{\rm out,0}^2 }	\right)^{\gamma-1}
					\left(
						\bar{R}_{\rm out}^2 - \frac{R_{\rm in,0}^2}{R_{\rm out,0}^2} \bar{R}_{\rm in}^2  
								\right)^{-\gamma}, 
	\label{eq:liner:Rout_bar} \\
		\frac{1}{2} \frac{\mathrm{d}^2 \bar{R}_{\rm in}}{\mathrm{d} \bar{t}^2}
			= & \,\Phi \left( \frac{R_{\rm out,0}}{R_{\rm in,0}} \right)^2  \bar{R}_{\rm in} \bar{p}_{\rm fuel}	 
					+ \Sigma
						 \left( \frac{R_{\rm out,0}}{R_{\rm in,0}} \right)^2  \bar{R}_{\rm in} \bar{B}_z^2
					- 2 \Lambda \bar{R}_{\rm in}
					\left( \frac{\widehat{m}/\rho_{\rm ref}}{\pi R_{\rm out,0}^2 }	\right)^{\gamma-1}
					\left( 
							 \bar{R}_{\rm out}^2 - \frac{R_{\rm in,0}^2}{R_{\rm out,0}^2} \bar{R}_{\rm in}^2 
							  \right)^{-\gamma}.
	\label{eq:liner:Rin_bar}
\end{align}
\end{widetext}
In this two-layer liner model, there are four important dimensionless parameters, which represent ratios of the various forms of energy relevant to MagLIF implosions:
\begin{align}
	\Pi & 	\doteq \frac{\mu_0 I_\star^2 }{4 \pi \widehat{m} R_{\rm out,0}^2 /  t_\varphi^2}
				\sim \frac{\rm Magnetic~potential~energy}{\rm Liner~kinetic~energy }, \\
	\Phi & 	\doteq \frac{4}{3} \frac{E_{\rm preheat}  }{\widehat{m} h R_{\rm out,0}^2 /  t_\varphi^2}
				\sim  \frac{\rm Preheat~energy}{\rm Liner~kinetic~energy },\\
	\Lambda & 	\doteq \frac{p_{\rm ref}/ \rho_{\rm ref}}{ R_{\rm out,0}^2 /   t_\varphi^2 }
						\sim  \frac{\rm Liner~internal~energy}{\rm Liner~kinetic~energy }, 
\end{align}
\begin{align}
	\Sigma 		& 	\doteq \frac{\pi}{\mu_0} \frac{B_{z,0}^2 R_{\rm in,0}^2}{ \widehat{m} R_{\rm out,0}^2 /   t_\varphi^2 }
						\sim  \frac{\rm Axial~magnetic~energy}{\rm Liner~kinetic~energy }.
\end{align}
The parameter $\Pi$ represents a ratio of the characteristic magnetic potential energy to the liner kinetic energy.  It denotes how strongly the magnetic drive accelerates the liner.\cite{Ryutov:2014hr}  Larger $\Pi$ values are associated with higher currents delivered to the load, longer timescales $t_\varphi$, less massive liners, and/or more compact liners.  Along with the parameter $c_6$ in \Eq{eq:electrical:c6}, $\Pi$ characterizes the nonlinear dynamical coupling between the liner and the circuit.  The dimensionless parameter $\Phi$ measures the relative importance of the fuel internal energy to the liner kinetic energy.  It characterizes the pushback done by the fuel on the imploding liner.\cite{Schmit:2020jd}  To increase $\Phi$ relative to $\Pi$, one clearly needs to increase the preheat per-unit-length
\begin{equation}
	\widehat{E}_{\rm preheat} \doteq E_{\rm preheat}/ h.
\end{equation}
The dimensionless quantity $\Lambda$ characterizes the relative importance of the liner internal energy and is absent in the thin-shell model.  This parameter is proportional to the ratio squared of the characteristic sound velocity $\sqrt{\gamma p_{\rm ref}/\rho_{\rm ref} }$ inside the liner and the characteristic implosion velocity $(\sim R_{\rm out,0}/t_\varphi)$ of the liner.  Finally, the last parameter $\Sigma$ denotes the ratio of the energy associated with the axial magnetic field within the fuel and the liner kinetic energy. This parameter measures the amount of pushback of the flux-compressed axial magnetic field onto the liner.  It is worth noting that the ratio $\Phi/\Sigma$ is effectively the ratio of the fuel pressure and the axial magnetic-field pressure, \ie the well-known $\beta$ parameter.  In convenient units, these parameters are written as
\begin{align}
	\Pi = & \, (10)^{-8} \frac{[I_\star({\rm MA})]^2 \cdot [t_{\rm \varphi} ({\rm ns})]^2 }
											{[\widehat{m} ({\rm g/cm})] 
												\cdot [R_{\rm out,0}({\rm cm})]^2} ,
		\label{eq:liner:Pi}		\\
	\Phi = & \, \frac{4}{3} \cdot (10)^{-8} 
						\frac{ [E_{\rm preheat} ({\rm kJ})] \cdot [t_{\rm \varphi} ({\rm ns})]^2 }
									{[\widehat{m} ({\rm g/cm})] \cdot [h ({\rm cm})] \cdot [R_{\rm out,0}({\rm cm})]^2}	,
		\label{eq:liner:Phi}
\end{align}

\begin{align}
	\Lambda = & \,    (10)^{-6}		\frac{[p_{\rm ref}({\rm Mbar})] \cdot [t_{\rm \varphi} ({\rm ns})]^2 }
													{ [\rho_{\rm ref}({\rm g/cm}^3)] \cdot [R_{\rm out,0}({\rm cm})]^2 }, 
		\label{eq:liner:Lambda}	\\
	\Sigma =  	& \,   \frac{5}{2} \cdot (10)^{-11}
										\frac{[B_{z,0}({\rm T})]^2 \cdot [R_{\rm in,0}({\rm cm})]^2 
													\cdot [t_{\rm \varphi} ({\rm ns})]^2 }
													{[\widehat{m} ({\rm g/cm})] \cdot [R_{\rm out,0}({\rm cm})]^2 }.
		\label{eq:liner:Sigma}		
\end{align}

There are two other dimensionless factors appearing in \Eqs{eq:liner:Rout_bar} and \eq{eq:liner:Rin_bar} that are of less importance.  For high-AR liners, the dimensionless factor $\widehat{m}/(\pi R_{\rm out,0}^2\rho_{\rm ref} )$ is approximately written as
\begin{equation}
	\frac{\widehat{m}/\rho_{\rm ref}}{\pi R_{\rm out,0}^2} 
		\simeq \frac{\rho_{\rm liner,0}}{\rho_{\rm ref}} \frac{2}{\rm AR } .
	\label{eq:liner:factor}
\end{equation}
Thus, it is a function of the material properties of the liner and its initial aspect ratio.  The second dimensionless factor $(R_{\rm in,0}/R_{\rm out,0})^2$ depends on the liner geometry and can be rewritten as 
\begin{equation}
	\left( \frac{R_{\rm in,0}}{R_{\rm out,0}} \right)^2 \simeq 1-\frac{2}{\rm AR}
	\label{eq:liner:factor2}
\end{equation}
for high-AR liners.
 
The curious reader might wonder about the origins of the specific interpretation given to the parameters in \Eqs{eq:liner:Pi}--\eq{eq:liner:Sigma} and ask why the parameters in \Eqs{eq:liner:Pi}--\eq{eq:liner:Sigma} are more important than those in \Eqs{eq:liner:factor} and \eq{eq:liner:factor2}.  We can answer these questions following an energy-conservation argument.  Let us multiply \Eq{eq:liner:Rout_bar} by $\smash{\mathrm{d}\bar{R}_{\rm out}/ \mathrm{d}\bar{t}}$ and \Eq{eq:liner:Rin_bar} by $\smash{(R_{\rm in,0}/R_{\rm out,0})^2 \mathrm{d}\bar{R}_{\rm in}/ \mathrm{d}\bar{t}}$.  Summing the resulting equations and integrating leads to an energy-conservation equation:
\begin{widetext}
\begin{align}
	\underbrace{-\Pi \int \frac{\bar{I}_l^2}{\bar{R}_{\rm out}}
					 \, \mathrm{d}\bar{R}_{\rm out}}_{\rm External~magnetic~energy~extracted}
		=  & \, 	\underbrace{
						\frac{1}{4} \left( \frac{R_{\rm in,0}}{R_{\rm out,0}} \right)^2 
							\left( \frac{\mathrm{d} \bar{R}_{\rm in}}{\mathrm{d}\bar{t}}\right)^2
						+\frac{1}{4} \left( \frac{\mathrm{d} \bar{R}_{\rm out}}{\mathrm{d}\bar{t}}\right)^2
					}_{\rm Liner~kinetic~energy}
				\quad 
				\underbrace{
					- \frac{\Phi}{2} \int  \bar{p}_{\rm fuel}	 \, \mathrm{d} \bar{R}_{\rm in}^2
				}_{\rm Work~on~fuel~column}
				\notag \\
			&	\underbrace{
				- \frac{\Sigma}{2}  
						 \int  \bar{B}_z^2 \, \mathrm{d} \bar{R}_{\rm in}^2
				}_{\mathrm{Work~on~flux~compressed}~B_z}
				+ 
				\underbrace{
						\frac{\Lambda}{\gamma-1}
					\left[ \left( \frac{p_{\rm liner}}{p_{\rm ref} } \right)^{(\gamma-1)/\gamma}
							-	\left( \frac{p_{\rm liner,0}}{p_{\rm ref} } \right)^{(\gamma-1)/\gamma}
					\right]
				}_{\rm Work~on~compressed~liner},
	\label{eq:liner:integration}
\end{align}
\end{widetext}
where we substituted \Eq{eq:liner:pliner}. We have explicitly denoted the physical meaning of each term in \Eq{eq:liner:integration}.  The quantity $p_{\rm liner,0}$ denotes the initial pressure within the liner, which is negligible.  From \Eq{eq:liner:integration}, we conclude that the available magnetic energy extracted by the imploding liner is converted to liner kinetic energy, compression of the fuel column, flux-compression of the axial magnetic field, and compression of the liner itself.  The main energy channels appearing in \Eq{eq:liner:integration} are characterized by the four parameters given in \Eqs{eq:liner:Pi}--\eq{eq:liner:Sigma}.  

%%%%%%%%%%%%%%%%%%%%%%%%%%%%%%%%%%%%%%%%%%%%%%%%%
%%%%%%%%%%%%%%%%%%%%%%%%%%%%%%%%%%%%%%%%%%%%%%%%%
\section{Liner stability and robustness to MRT instabilities}
\label{sec:stability}

{\color{black}MagLIF implosions are susceptible to the magneto-Raleigh--Taylor (MRT) instability.\cite{Harris:1962hu,Weis:2015hk,Velikovich:2015jl,Sinars:2010de,McBride:2012db,McBride:2013gda,Awe:2014gba,Ruiz:2022aa}}  As discussed in \Refa{Schmit:2020jd}, the MRT modes that most decrease the amount of pdV work done on the fuel and reduce confinement are those that maximize the perturbations at the liner interior surface.  When accounting for MRT feedthrough, one can estimate the number of $e$-foldings of the most dangerous MRT modes during the acceleration phase of the implosion:\cite{Schmit:2020jd}

\begin{equation}
	\Gamma_{\rm max}(t)
			\simeq	\frac{1}{4}\frac{ \mathrm{IFAR}(t)}{ R_{\rm out}(t)}  
						\left( \int_0^t \sqrt{|\ddot{R}_{\rm out}|} \, \mathrm{d} t' \right)^2,
	\label{eq:stability:Gamma}
\end{equation}
where the liner in-flight aspect ratio (IFAR) is defined as the ratio of the imploding shell outer radius and the shell thickness:
\begin{equation}
	\mathrm{IFAR}(t) 
		\doteq \frac{R_{\rm out}(t)}{\delta R(t)}
		= \frac{R_{\rm out}(t)}{R_{\rm out}(t) - R_{\rm in}(t)} .
\end{equation}
Notably, $\mathrm{IFAR}(0) = \mathrm{AR}$, where AR is the initial liner aspect ratio.  The IFAR is a commonly used metric in the ICF community to measure the stability of imploding shells.\cite{Bose:2017jf}  In general, larger IFAR values indicate that the shells are more unstable to Rayleigh--Taylor instabilities.

{\color{black}Equation~\eq{eq:stability:Gamma} serves as our measure of the susceptibility of the liner towards the MRT instability; when $\Gamma_{\rm max}(t)$ is large, MRT perturbations grow to larger amplitudes and can  compromise the integrity of the liner.}  To maintain (or even improve) liner stability of MagLIF implosions, one can \textit{simultaneously} satisfy the following three conditions.  {\color{black}First, one can seek implosion configurations where the outer convergence ratio $\mathrm{CR}_{\rm out}(t) \doteq R_{\rm out,0}/R_{\rm out}(t)$ of the liner is maintained or reduced.}  Second, the characteristic amplitude of the liner IFAR should not increase when changing the MagLIF load parameters.  Third, the integral of the square root of the liner acceleration history should not increase when varying the liner parameters. 

The liner IFAR can be calculated numerically when solving \Eqs{eq:liner:Rout} and \eq{eq:liner:Rin}.  However, one can derive an approximate expression for the IFAR as follows. Subtracting \Eqs{eq:liner:Rout} and \eq{eq:liner:Rin} gives
\begin{align}
	\frac{\widehat{m}}{2}	\, \frac{\mathrm{d}^2 \delta R}{\mathrm{d} t^2} 
			=&-2\pi \left[ R_{\rm in}  p_{\rm fuel}
				+  R_{\rm in}  p_{\rm mag,int} 
				+ R_{\rm out} p_{\rm mag,ext} \right.
				\notag \\
			  & \left.
				- (R_{\rm out} + R_{\rm in} ) \,  p_{\rm liner} \right] .
	\label{eq:stability:IFAR_aux}
\end{align}
We assume that the left-hand side of \Eq{eq:stability:IFAR_aux} is negligible so that the liner pressure is in a quasi-equilibrium state.  This effectively eliminates the high-frequency oscillations of the liner as it vibrates during the implosion.  We then substitute $R_{\rm in}=R_{\rm out} - \delta R$ into \Eq{eq:stability:IFAR_aux}.  In the thin-shell limit, we neglect $\delta R/R_{\rm out}$ terms on the right-hand side.  As previously mentioned, this limit is valid as long as the liner has not converged too much.  In the regime of validity of this approximation, the fuel pressure and internal magnetic pressure are small compared to the external magnetic pressure driving the implosion.  Therefore, we may drop those terms.  Next, we substitute \Eq{eq:liner:pliner} for the internal pressure of the liner and solve the resulting algebraic equation for $\delta R$.  This gives
\begin{equation}
	\mathrm{IFAR}(t)
		\simeq	2 \pi \,
					\frac{R_{\rm out}^2(t) \rho_{\rm ref}}{\widehat{m} }  
					\left( \frac{p_{\rm mag,ext}(t) }{ 2p_{\rm ref} } \right)^{1/\gamma} .
	\label{eq:stability:IFAR}
\end{equation}
%
%Note that this factor also appears in \Eq{eq:stability:IFAR} for the IFAR.   eq:liner:factor}

Equation \eq{eq:stability:IFAR} is in general agreement with the results derived in \Refs{Schmit:2020jd,Nora:2014iq}.  In those works, a model for the IFAR was derived from a fluid-type momentum conservation equation for an imploding thin shell.  Since the starting point here is the two-layer liner model, there are discrepancies with the numerical coefficients, as well as the dependency on the ratio $p_{\rm fuel}/p_{\rm mag}$.  However, more importantly for this work, the derived expression for the IFAR is proportional to $\smash{R_{\rm out}^2p_{\rm mag,ext}^{1/\gamma}/\widehat{m}}$ as in \Refs{Schmit:2020jd,Nora:2014iq}.

%%Another remark is that the \Eq{eq:stability:IFAR} does not give the initial AR of the liner in the limit when the external magnetic pressure is null.  To fix this problem, one should subtract from \Eq{eq:liner:pliner} the initial internal pressure of the liner.  Following the same derivation above, one obtains
%%
%\begin{equation}
%	\mathrm{IFAR}(t)
%		\simeq	2 \pi \,
%					\frac{R_{\rm out}^2(t) \rho_{\rm ref}}{\widehat{m} }  
%					\left[ \frac{p_{\rm mag}(t) }{ 2p_{\rm ref} } 	
%							+ \left( \frac{\rho_{\rm liner,0}}{\rho_{\rm ref}} \right)^\gamma \right]^{1/\gamma} .
%	\label{eq:basic:IFARII}
%\end{equation}
%%
%%The second term inside the brackets is necessary to have IFAR(0) = AR but rapidly becomes negligible once the external magnetic pressure increases.

In terms of dimensionless quantities, the number of $e$-foldings $\Gamma_{\rm max}$ in \Eq{eq:stability:Gamma}  is written as
\begin{equation}
	\Gamma_{\rm max} (\bar{t})
			=	\frac{1}{4} \frac{\mathrm{IFAR}(\bar{t})}{\bar{R}_{\rm out}(\bar{t})}
				\left( 
					\int^{\bar{t}}_0 \sqrt{|\ddot{\bar{R}}_{\rm out}(\bar{t}')|} \, \mathrm{d}\bar{t}' 				
				\right)^2.
	\label{eq:dim:Gamma}
\end{equation}
We can also rewrite the liner IFAR in terms of dimensionless variables as follows:
\begin{gather}
	\mathrm{IFAR}(\bar{t}) 
		=	\Psi \, \bar{R}_{\rm out}^2 \, \left( \frac{\bar{I}_l}{\bar{R}_{\rm out}} \right)^{2/\gamma} ,
	\label{eq:dim:IFAR}
\end{gather}
where the dimensionless parameter $\Psi$ represents the characteristic amplitude of the liner IFAR:
\begin{align}	
	\Psi	\doteq 
		 \, 2 \pi \, 
			\frac{		[R_{\rm out, 0}({\rm cm})]^{2-2/\gamma} 
						\cdot [\rho_{\rm ref}({\rm g/cm^3})] 
						\cdot [I_\star ({\rm MA})]^{2/\gamma}
			}{		({\color{black}400}\pi)^{1/\gamma} 
					\cdot  [\widehat{m}({\rm g/cm})] 
					\cdot [p_{\rm ref}({\rm Mbar})]^{1/\gamma} 
			}.
	\label{eq:dim:Psi}
\end{align}
Three factors contribute to the characteristic amplitude $\Psi$ of the IFAR.  First, $\Psi$ depends on the initial geometry of the liner:  the term {\color{black}$R_{\rm out,0}^2/\widehat{m}$} is proportional to the initial aspect ratio (AR) of the liner.  For this reason, the liner initial AR is often considered as a measure of liner robustness towards instabilities.\citep{Slutz:2010hd}  Second, $\Psi$ depends on the characteristic magnetic pressure acting on the liner, which is proportional to $(I_\star/R_{\rm out,0})^2$.  Higher magnetic pressures lead to larger values  of the IFAR due to enhanced magnetic compression.  Finally, $\Psi$ depends on the compressibility of the liner, which manifests itself via the parameters $\rho_{\rm ref}$, $p_{\rm ref}$, and $\gamma$.

%%%%%%%%%%%%%%%%%%%%%%%%%%%%%%%%%%%%%%%%%%%%%%%%%
%%%%%%%%%%%%%%%%%%%%%%%%%%%%%%%%%%%%%%%%%%%%%%%%%
\section{Fuel energetics}
\label{sec:energy}

In a simplified manner, the energy-balance equation for the fuel is given by\cite{Slutz:2010hd,McBride:2015ga}
\begin{equation}
		\frac{\mathrm{d} U}{\mathrm{d} t}	
						=	 P_{\rm preheat} 
							+ P_{\rm pdV} 
							+ P_{\rm \alpha} 
							- P_{\rm rad} 
							- P_{\rm c}
							- P_{\rm end} 
							+ P_{\rm other}.
	\label{eq:energy:U}
\end{equation}
The right-hand side of \Eq{eq:energy:U} includes the various energy source and sink mechanisms present in a MagLIF implosion.  $\smash{P_{\rm preheat}(t)}$ is the external fuel preheat rate, $\smash{P_{\rm pdV}(t)}$ is the pdV-work rate on the fuel, and $\smash{P_{\alpha}(t)}$ is heating rate due to $\alpha$ particles.  Regarding the energy-loss mechanisms, $\smash{P_{\rm rad}(t)}$ is the radiative cooling rate, $\smash{P_{\rm c}(t)}$ is the cooling rate due to conduction losses, and $\smash{P_{\rm end}(t)}$ represents an energy-loss rate due to mass flow through the open liner ends.  Finally, $P_{\rm other}(t)$ represents other energy source and loss mechanisms that are not discussed in this work;  for example, additional radiation losses due to mixing of contaminants into the fuel $(P_{\rm other} = - P_{\rm rad, mix})$,\cite{Knapp:2019gf} degradation of the compressional pdV work caused by MRT instabilities $(P_{\rm other} = - P_{\rm MRT})$,\cite{Bose:2017jf} and recirculation of energy from ablation of the liner material caused by local heat fluxes {\color{black}$(P_{\rm other} = P_{\rm ablation})$}.\cite{Hurricane:2019hd}  In the remainder of this section, we shall discuss in further detail the terms appearing in \Eq{eq:energy:U}.

%%%%%%%%%%%%%%%%%%%%%%%%%%%%%%%%%%%%%%%%%%%%%%%%%
\subsection{Energy source and sink mechanisms}
\label{sec:energy:sources_sinks}

\msection{Fuel preheat} Before external preheat occurs, the internal gas pressure is negligible compared to the compressive magnetic pressure of the pulsed-power driver.  Therefore, it is a good approximation to neglect the fuel internal energy prior to the laser preheat.  Once laser preheat occurs, we assume that an energy $\smash{E_{\rm preheat}}$ is instantaneously deposited uniformly into the fuel volume.\cite{foot:preheat} Within this approximation, we model the preheat as follows:
\begin{equation}
	P_{\rm preheat}
			\doteq E_{\rm preheat} \,
			\delta(t- t_{\rm preheat}),
	\label{eq:basic:P_preheat}
\end{equation}
where $t_{\rm preheat} $ is the time at which preheat occurs.

\msection{Compressional pdV work} After preheat, the imploding liner performs work onto the fuel gas in its interior.  The pdV work rate is given by
%
%\begin{subequations}
\begin{equation}
	P_{\rm pdV} 
				\doteq - p_{\rm fuel}(t) \frac{\mathrm{d}}{\mathrm{d}t}\pi R_{\rm in}^2 h
				=	-	\frac{4}{3}  \frac{U}{R_{\rm in}} \frac{\mathrm{d} R_{\rm in} }{\mathrm{d}t},
	\label{eq:basic:P_PdV}
\end{equation}
where we substituted \Eq{eq:liner:pfuel}.

\msection{Alpha heating} In a self-heating scenario, the energetic $\alpha$ particles created by DT fusion reactions can be confined by the axial magnetic field $B_z$ within the plasma.  As the $\alpha$ particles collide with the fuel electrons, they lose their energy and heat the background fuel plasma.  A simple expression for the $\alpha$-heating rate is the following:\cite{foot:alpha}
\begin{equation}
	P_\alpha = \varepsilon_\alpha \frac{\mathrm{d} Y}{\mathrm{d}t} \eta_\alpha,
	\label{eq:basic:P_alpha}
\end{equation}
where $\varepsilon_\alpha = 3.5$~MeV is the energy of the $\alpha$ particles and $\mathrm{d}Y/\mathrm{d}t$ is the neutron-yield rate [see \Eq{eq:basic:yield_rate}].  The parameter $\eta_\alpha(t)\leq 1$ represents the fraction of $\alpha$ particles trapped within the fuel column.  When assuming that $\alpha$ particles are primarily lost radially instead of axially, we may approximate the fraction $\eta_\alpha$ by\cite{Basko:2002dv}
\begin{equation}
	\eta_\alpha \simeq \frac{x_\alpha + x_\alpha^2}{1+13x_\alpha/9 + x_\alpha^2},
	\label{eq:basic:eta_alpha}
\end{equation}
where the dimensionless parameter $x_\alpha$ is defined as
\begin{equation}
	x_\alpha \doteq 
						\frac{8}{3} 
						\left( \frac{R_{\rm in}}{\ell_\alpha} + 
								\frac{(R_{\rm in}/\varrho_\alpha)^2}{\sqrt{9(R_{\rm in}/\varrho_\alpha)^2 + 1000}} \right).
	\label{eq:basic:x_alpha}
\end{equation}
The parameter $x_\alpha$ depends on two dimensionless quantities.  First, $R_{\rm in}/\ell_\alpha$ is the ratio of the fuel-column radius and the stopping length of the 3.5-MeV $\alpha$ particles.  In terms of practical units, this dimensionless parameter is\cite{Atzeni:2009phys}
\begin{equation}
	\frac{R_{\rm in}}{\ell_\alpha}
		 = 9.34  \, \frac{\ln \Lambda_{\alpha e} \cdot [\rho R_{\rm in} (\mathrm{g/cm^2})]  }{[T(\mathrm{keV})]^{3/2}},
	\label{eq:basic:a}
\end{equation}
where $\ln \Lambda_{\alpha e}$ is the Coulomb logarithm for collisions between $\alpha$ particles and electrons.  The parameter $R_{\rm in}/\varrho_\alpha$ is the ratio of the fuel-column radius $R_{\rm in}$ to the Larmor radius $\varrho_\alpha\doteq v_\alpha/\Omega_\alpha$ of the 3.5-MeV $\alpha$ particles, where $v_\alpha\doteq \sqrt{2 m_\alpha \varepsilon_\alpha}$ is their initial velocity, $\Omega_\alpha \doteq 2e B_z / m_\alpha$ is the gyrofrequency of $\alpha$ particles, $e$ is the elementary charge, and $m_\alpha$ is the mass of an $\alpha$ particle.  In convenient units, $R_{\rm in}/\varrho_\alpha$ can be written as
\begin{equation}
	\frac{R_{\rm in}}{\varrho_\alpha}
		 \doteq \frac{ [B_z(\mathrm{T})] \cdot [R_{\rm in}(\mathrm{cm})] }{26.5}.
	\label{eq:basic:b}
\end{equation}
When $R_{\rm in}/\varrho_\alpha\gg1$, $\alpha$ particles are well confined within the fuel plasma column.  {{\color{black}Note that the magnetic-field--radius product $\langle B_z R_{\rm in}\rangle$ is a performance metric often inferred in MagLIF experiments.\cite{Schmit:2014fg,Knapp:2015kc,Lewis:2021kz}}  In MagLIF implosions, $R_{\rm in}/\varrho_\alpha \gg R_{\rm in}/\ell_\alpha$, so magnetic confinement is the main mechanism for trapping $\alpha$ particles, not collisions.

\msection{Radiation losses} Electron Bremsstrahlung emission is the dominant mechanism for radiation losses.  When assuming that the fuel plasma is optically thin and that any radiation energy deposited into the MagLIF liner does not alter the implosion dynamics, the approximate energy-loss rate of a hot, uniform, cylindrical plasma is\cite{Atzeni:2009phys} 
\begin{equation}
	P_{\rm rad}
		\doteq  \frac{64}{3 \sqrt{2 \pi}} 
					\left( \frac{e^2}{4\pi \epsilon_0} \right)^3 \frac{1}{m_e c^2 \hbar}
					\sqrt{\frac{k_B T}{m_e c^2}} \frac{Z^3 \rho^2}{m_i^2}
					\pi R_{\rm in}^2 h,
\end{equation}
where $\epsilon_0$ is the vacuum permittivity, $m_e$ is the electron mass, $\hbar\doteq h/(2\pi)$ is the Planck constant, and $c$ is the speed of light.  The factor $\pi R_{\rm in}^2 h$ denotes the volume of the plasma column.  Here we do not consider other sources of radiation emission such as those coming from impurity mixing in MagLIF.\cite{Knapp:2019gf}  Considering the effects of additional sources and their scaling is left for future work.

{\color{black}
\msection{Conduction losses} Due to the radial compression of z-pinch plasmas, thermal conduction losses will be dominated by thermal flux in the radial direction.  When an isobaric hot plasma comes in contact with a cold liner wall, an advective flow arises from the center of the hot plasma core to the walls in order to maintain pressure equilibrium.  The advective flow can carry the heat flux faster than thermal diffusion, specially in the high-magnetization regime $x_e\gg1$.\cite{Vekshtein:1983aa,Vekshtein:1986aa}  Here $x_e\doteq \Omega_e \tau_e$ is the electron Hall parameter, $\Omega_e\doteq e B_z/m_e$ is the electron gyrofrequency,
\begin{align}
	\tau_e	&	\doteq 	\frac{3(4\pi \varepsilon_0)^2 m_i \sqrt{m_e} (k_B T)^{3/2}}
						{4\sqrt{2\pi} Z^2 \rho \, e^4 \ln \Lambda },
						\label{eq:basic:tau_e}	
\end{align}
is the electron--ion collision time, and $\ln \Lambda$ is the Coulomb logarithm.  In convenient units, the electron Hall parameter can be written as
\begin{equation}
	x_e  = 3.20	\cdot (10)^{-3}
				 \frac{A \cdot \left[T({\rm keV})\right]^{3/2}\cdot\left[B_z({\rm T})\right]}
				 		{\ln \Lambda \cdot \left[\rho({\rm g/cm^3})\right]}
					\label{eq:energy:xe}	\\
\end{equation}
for hydrogen-like plasma.  Due to internal advection flows, the loss rates of thermal energy from the hot plasma core can be higher than the estimates based on the electron and ion thermal conductivities with magnetization effects included.\cite{Braginskii:1965vl}

In \Refa{Velikovich:2015gs}, the one-dimensional problem of a semi-infinite, magnetized hot plasma in contact with a cold liner wall was studied.  The liner wall was modeled as an ideal heat sink.  Using self-similar solutions of the classical, collisional, Braginskii, plasma-transport equations, it was found that the effective thermal diffusivity $\chi_{\rm \perp,eff}$ perpendicular to the magnetic field scales as $x_e^{-1}$ so that
\begin{equation}
	\chi_{\rm \perp ,eff} \simeq 32 D_B, 
	\label{eq:energy:chiperp}
\end{equation}
where $D_B$ is the well-known Bohm-diffusion coefficient
\begin{equation}
	D_B \doteq \frac{1}{16} \frac{k_B T_e}{m_e \Omega_e}.
	\label{eq:energy:DB}
\end{equation}
Hence, thermal losses in the highly magnetized regime follow a Bohm-like scaling.  This approximate result remains valid when ions in the hot plasma core are magnetized so that $x_i>1$ and $x_e>30.3 A^{1/2}$.  Here $x_i$ is the ion Hall parameter which relates to the electron Hall parameter by $x_i = \sqrt{2m_e/m_i} \, x_e \simeq 0.033 x_e/A^{1/2}$ for hydrogen-like plasma.

Based on the results in \Refa{Velikovich:2015gs}, the energy-loss rate $P_{\rm c}$ due to thermal conduction losses can be approximately modeled as follows:
\begin{align}
	P_{\rm c} 
		&	\doteq 2 \pi R_{\rm in} h \kappa_{\rm \perp,eff} \, \pd_r (k_B T) \notag \\
		&	\simeq 2\pi  h  \,\kappa_{\rm \perp,eff} \, k_B T, \notag \\
		&	= 12 \pi h \frac{\rho}{A m_p} \frac{(k_B T)^2}{m_e \Omega_e},
	\label{eq:energy:Pc}
\end{align}
where $\pd_r T \simeq T/R_{\rm in}$ and $\kappa_{\rm \perp,eff}=3(\rho/m_i) \chi_{\rm \perp ,eff}$ is the effective thermal conductivity.

In the following, we shall use \Eq{eq:energy:Pc} as our model to describe thermal conduction losses.  However, we note that the effective thermal diffusivity in \Eq{eq:energy:chiperp} is a result of an idealized picture transporting energy from a semi-infinite hot plasma to a cold wall.  Of course, the fuel inside MagLIF is imploding and compressibly heating in a cylindrical geometry.  Such effects are not included in \Refa{Velikovich:2015gs}.  Moreover, performance of MagLIF, \eg, the fusion yield, is sensitive to the specific profile of the fuel temperature, which is more sensitive to the specific dynamics of ion-conduction losses.
}

\msection{End losses} Cylindrical MagLIF liners contain openings at their ends to inject fuel and preheat energy.  Once the implosion is underway, hot fuel can leave the liner interior through these openings and take energy along with it.  Following \Refs{Slutz:2010hd,McBride:2015ga}, we model the escape of fuel across the top and bottom openings as a rarefraction wave propagating to the interior of the fuel.  According to \App{app:rarefraction}, the energy end losses are given by
\begin{equation}
	P_{\rm end}
		\simeq	0.90
				\frac{U c_s R_{\rm c}^2}{R_{\rm in}^2 h}
		=	2.71 \pi \frac{\rho }{m_i}k_B T
			 c_s R_{\rm c}^2 ,
		\label{eq:basic:P_end}
\end{equation}
where
\begin{equation}
	c_s \doteq \sqrt{2 \gamma_{\rm fuel} \, \frac{ k_B T }{m_i}}
\end{equation}
is the hydrodynamic (ion) sound speed and $\gamma_{\rm fuel}=5/3$.  Also, $R_{\rm c}(t)$ denotes the radius of the opening through which the fuel escapes the fluid.  It is given by the minimum between the radius $R_{\rm LEH}$ of the laser-entrance-hole (LEH) window\cite{HarveyThompson:2018dd} and the liner inner radius.  More specifically, $R_c(t) \doteq \min \{ R_{\rm LEH}, R_{\rm in}(t) \}$. 

%%%%%%%%%%%%%%%%%%%%%%%%%%%%%%%%%%%%%%%%%%%%%%%%%
\subsection{Generalized equation-of-state for the fusion fuel}
\label{sec:energy:EOS}

One can rewrite the differential equation \eq{eq:energy:U} as an integral equation that makes more explicit the effects of energy sources and sinks.  Substituting \Eq{eq:basic:P_preheat} and \eq{eq:basic:P_PdV} into \Eq{eq:energy:U} and integrating in time gives the following:
\begin{multline}
	\frac{U R_{\rm in}^{4/3}}{E_{\rm preheat} R_{\rm in,preheat}^{4/3}} \\
		=	\exp \left( \int^t_{t_{\rm preheat}} 
						\frac{P_\alpha - P_{\rm rad} - P_c - P_{\rm end}}{U} \, \mathrm{d}t'
					\right),
	\label{eq:basic:EOS_fuel}
\end{multline}
where $R_{\rm in,preheat} \doteq R_{\rm in}(t_{\rm preheat})$ is the liner inner radius at the moment of preheat.  To maximize pdV work, typically $R_{\rm in,preheat}\simeq R_{\rm in,0}$.

Equation \eq{eq:basic:EOS_fuel} serves as a generalized EOS for the fuel internal energy.  The form of this equation is reminiscent to that of previous works aimed at understanding the effects of $\alpha$ heating and of hydrodynamical instabilities on fuel compression.\cite{Hurricane:2019hd,Hurricane:2021cx,Bose:2017jf}  The interpretation of \Eq{eq:basic:EOS_fuel} is simple: in the absence of energy sources or sinks, the left-hand side of \Eq{eq:basic:EOS_fuel} represents perfect adiabatic compression of the fuel.  The exponential on the right-hand side represents deviations from ideal adiabatic compression, \ie changes in the entropy of the fuel.\cite{Hurricane:2019hd}  When the argument of the exponential is positive (\ie in the case of robust $\alpha$ heating), the internal energy of the fuel is larger than the purely adiabatic result at fixed inner convergence ratio $\mathrm{CR}_{\rm in } \doteq R_{\rm in,0} / R_{\rm in}$.  In the opposite case, when loss mechanisms are dominant, the fuel compression becomes less efficient:  to achieve the same fuel internal energy at fixed $E_{\rm preheat}$, the required inner convergence ratio must be larger.

The energy-source term corresponding to $\alpha$ heating in \Eq{eq:basic:EOS_fuel} is related to the general Lawson-criterion parameter $\chi$ for steady-state thermonuclear ignition that is often found in the ICF literature:\cite{Betti:2010fc}
\begin{align}
	\chi
		&	\doteq	\int^t_{t_{\rm preheat}} 	\frac{P_\alpha}{U} \, \mathrm{d}t'
			\notag \\
		&	=	\int^t_{t_{\rm preheat}} 
					\frac{1}{6} \varepsilon_\alpha  n_i^2 
					\frac{\langle \sigma v \rangle}{p_{\rm fuel} }  \, \eta_\alpha \, \mathrm{d}t'
				\notag \\
		&	=	\int^t_{t_{\rm preheat}} 
					\frac{1}{24} \varepsilon_\alpha  p_{\rm fuel}
					\frac{\langle \sigma v \rangle}{(k_BT)^2 }  \, \eta_\alpha \, \mathrm{d}t'.
	\label{eq:energy:chi}
\end{align}
where we substituted \Eqs{eq:liner:pfuel}, \eq{eq:basic:P_alpha} and \eq{eq:basic:yield_rate}.  Of course, significant neutron-yield production occurs during a relatively small window of time, known as the burn-width time $\tau_{\rm bw}$, which is small compared to the implosion time of the liner.  (In practice, $\tau_{\rm bw}$ is defined as the full-width half-maximum of the neutron-production rate.)  In terms of convenient units, $\chi$ can be approximated as
\begin{equation}
	\chi = 0.09 [p_{\rm fuel}({\rm Gbar})] \cdot [\tau_{\rm bw}({\rm ns})]
	\frac{[ \langle \sigma v \rangle (10^{-18} {\rm cm^3/s})] }{[T ({\rm keV}) ]^2}
	\eta_\alpha.
	\label{eq:energy:chi_convenient}
\end{equation}
Note that other Lawson parameters appearing in the ICF literature invoke a spatial average of the $\langle \sigma v \rangle/T^2$ over the fuel volume.\cite{Betti:2010fc}  We do not have such spatial averaging due to the zero-dimensional nature of the model presented in this paper, but the expression in \Eq{eq:energy:chi_convenient} could be readily improved so that the spatial-averaging operation is taken into account.  It is also worth noting that, absent any energy-recycling mechanism\cite{Hurricane:2019hd,Hurricane:2021cx} potentially occurring from ablation of the liner inner surface, \Eq{eq:energy:chi_convenient} includes the fraction $\eta_\alpha\leq1$ of trapped $\alpha$ particles.  
%Therefore, \Eq{eq:energy:chi_convenient} could serve as a potential candidate for a Lawson parameter specific to magneto-inertial-fusion systems.

The generalized EOS \eq{eq:basic:EOS_fuel} for the fusion fuel also provides insights on how the energy gain and loss mechanisms behave as the MagLIF liner implodes.  For the following calculations in this section, we consider that the quantities describing the fuel and axial magnetic field are adiabatically compressed to leading order when neglecting non-ideal mechanisms.  In terms of the inner convergence ratio $\mathrm{CR}_{\rm in} \doteq R_{\rm in,0} / R_{\rm in} = 1/\bar{R}_{\rm in}$, the parameters describing the fuel column behave as follows: $\smash{R_{\rm in } \sim \mathrm{CR}_{\rm in }^{-1} }$, $\smash{ U \sim \mathrm{CR}_{\rm in }^{4/3} }$, $\smash{ \rho \sim \mathrm{CR}_{\rm in }^{2} }$, $\smash{ T \sim \mathrm{CR}_{\rm in }^{4/3} }$, and $\smash{ B_z \sim \mathrm{CR}_{\rm in }^{2} }$.  When substituting the power-law expression \eq{eq:basic:reactivity_simple} for the DT reactivity into $\chi$, we find that the relative energy-gain rate due to $\alpha$ heating grows as
\begin{equation}
	\frac{1}{\tau_{\rm E,\alpha}} 
		\doteq {\color{black}\frac{P_{\alpha}}{U}} 
		\sim  \mathrm{CR}_{\rm in }^{5.69},
\end{equation}
where $\tau_{\rm E,\alpha}$ is the characteristic energy-gain time due to $\alpha$ heating.  (In the calculation above, we ignored the effects of $\eta_\alpha$ which increases for larger CR$_{\rm in}$ values.)  As expected, $\alpha$ heating is a process that rapidly increases near peak compression.  

The relative radiation-loss rate behaves as
\begin{equation}
	\frac{1}{\tau_{\rm E,rad}} \doteq \frac{P_{\rm rad}}{U} \sim  \mathrm{CR}_{\rm in }^{4/3},
\end{equation}
where $\tau_{\rm E,rad}$ is a characteristic energy confinement time due to radiation losses alone.  Thus, relative energy losses due to radiation increase sharply as the implosion progresses towards stagnation.

{\color{black}The relative energy losses due to thermal conduction depend on the inner convergence ratio as
\begin{equation}
	\frac{1}{\tau_{\rm E,c}} 
		\doteq	\frac{P_c}{U}
		\sim \mathrm{CR}_{\rm in}^{4/3},
\end{equation}
where $\tau_{\rm E,c}$ is the characteristic energy confinement time associated to thermal-conduction losses.  Interestingly, thermal conduction losses have a similar dependency on $ \smash{\mathrm{CR}_{\rm in }^{4/3}}$ as radiation losses.}

%Depending on the magnetization regime of electrons and ions, we have the following dependencies for thermal-conduction losses:
%%
%\begin{gather}
%	\frac{1}{\tau_{\rm E,ce}} \doteq
%		\frac{P_{\rm ce}}{U}\sim\left\{
%                \begin{array}{ll}
%                  \mathrm{CR}_{\rm in}^{4/3},\qquad ~~\mathrm{marginally~magnetized} \\
%                  \mathrm{CR}_{\rm in}^{-2/3},\qquad \mathrm{strongly~magnetized}
%                \end{array}
%              \right.
%    \label{eq:energy:tau_Ece} \\
%	\frac{1}{\tau_{\rm E,ci}} \doteq
%		\frac{P_{\rm ci}}{U}\sim\left\{
%                \begin{array}{ll}
%                  \mathrm{CR}_{\rm in}^{10/3},\qquad \mathrm{unmagnetized} \\
%                  \mathrm{CR}_{\rm in}^{4/3},\qquad ~\mathrm{marginally~magnetized}
%                \end{array}
%              \right.
%     \label{eq:energy:tau_Eci}
%\end{gather}
%%
%where $\tau_{\rm E,ce}$ and $\tau_{\rm E,ci}$ are characteristic energy confinement times due to electron and ion conduction losses, respectively.  Here we used, $x_{e,i} \propto \mathrm{CR}_{\rm in}^2$.  As shown in \Eq{eq:energy:tau_Ece}, electron thermal-conduction losses increase momentarily after the fuel preheat occurs, but once electrons become sufficiently magnetized, these losses saturate or become weaker as the implosion progresses.  In contrast, ion thermal-conduction losses are expected to increase as the implosion progresses.

The characteristic energy-confinement time due to end losses depends on the inner convergence ratio as
\begin{equation}
	\frac{1}{\tau_{\rm E,end}} \doteq
		\frac{P_{\rm end}}{U} \sim  \mathrm{CR}_{\rm in }^{2/3}.
\end{equation}
The relative end-loss rate $1/\tau_{\rm E,end}$ climbs at a slower rate than its radiation and ion-conduction counterparts, which means that its effects on the degradation of the ideal compression of the fuel are more {\color{black}uniformly distributed} in time from the moment of preheat to stagnation.

Finally, the total energy confinement time due to gains and losses can be defined as
%
%\begin{equation}
%	\frac{1}{\tau_E} 
%		\doteq \frac{1}{\tau_{\rm E,rad}}
%					+	\frac{1}{\tau_{\rm E,ce}}
%					+	\frac{1}{\tau_{\rm E,ci}}
%					+	\frac{1}{\tau_{\rm E,end}}
%					- 	\frac{1}{\tau_{\rm E,\alpha}}.
%\end{equation}
\begin{equation}
	\frac{1}{\tau_E} 
		\doteq \frac{1}{\tau_{\rm E,rad}}
					{\color{black}+	\frac{1}{\tau_{\rm E,c}}}
					+	\frac{1}{\tau_{\rm E,end}}
					- 	\frac{1}{\tau_{\rm E,\alpha}}.
\end{equation}
Tracking the ratio of the burn width $\tau_{\rm bw}$ and $\tau_E$ (along with its individual components) can serve as a good {\color{black}measure} of how the energy gain and loss mechanisms vary when studying different MagLIF configurations.

As a final concluding remark for this section, we wrote \Eq{eq:basic:EOS_fuel} as a generalized EOS for the fuel internal energy $U$.  However, a similar expression can be derived for the fuel pressure $p_{\rm fuel}$.  Substituting \Eq{eq:liner:pfuel} into \Eq{eq:basic:EOS_fuel} leads to an equation similar to \Eq{eq:basic:EOS_fuel} with $U/E_{\rm preheat}$ replaced by $p_{\rm fuel}/p_{\rm preheat}$ and the exponent $4/3$ changed to $10/3$.  To obtain a generalized EOS for the fuel temperature, one can substitute \Eq{eq:liner:pfuel} into \Eq{eq:basic:EOS_fuel} and also use the result in \Eq{eq:mass:EOS} to account for fuel-mass losses.

%%%%%%%%%%%%%%%%%%%%%%%%%%%%%%%%%%%%%%%%%%%%%%%%%
\subsection{Dimensionless form}

We shall now write \Eq{eq:basic:EOS_fuel} in dimensionless form.  We introduce the following two dimensionless variables:
\begin{equation}
	\bar{\rho}  \doteq \frac{\rho}{\rho_0} , \qquad
	\bar{T}  \doteq \frac{T}{T_{\rm preheat} } ,
	\label{eq:energy:variables3}
\end{equation}
where $\rho_0\doteq \rho(0)$ is the initial fuel mass density and
\begin{equation}
	k_B T_{\rm preheat} 
		  \doteq \frac{p_{\rm preheat}}{2 (\rho_0/m_i)}
		  \label{eq:energy:T_preheat}	
\end{equation}
is the characteristic temperature of the fuel achieved after the preheat stage.
%
%equivalent mean fuel temperature after preheat assuming that the preheat energy was deposited before the liner began converging.  In convenient units, 
%
%\begin{equation}
%	T_{\rm preheat} ({\rm keV}) 
%		  = 2.77\cdot (10)^{-6} \, \frac{[\widehat{E}_{\rm preheat} (\mathrm{kJ/cm})]  }
%									{\left[\rho_0 ({\rm g/cm^3})\right] \cdot [R_{\rm in,0}({\rm cm})]^2 }.
%\end{equation}
%
When substituting \Eqs{eq:liner:pfuel}, \eq{eq:liner:variables2} and \eq{eq:energy:variables3} into \Eq{eq:basic:EOS_fuel}, we obtain
\begin{widetext}
%\begin{equation}
%	\bar{U}
%		=	\frac{\mathrm{CR}_{\rm in}^{4/3}}{\mathrm{CR}_{\rm in,preheat}^{4/3} }
%			 \exp \left[ \int^{\bar{t}}_{\bar{t}_{\rm preheat}}  \left(
%						 \Upsilon_\alpha 
%						 			\, \bar{\rho} \, \bar{T}^{2.77}
%						 			\eta_\alpha
%						- \Upsilon_{\rm rad} \frac{\bar{\rho}}{\bar{T}^{1/2}} 
%						- \Upsilon_{\rm ce} \frac{ \bar{T}^{5/2} }{ \bar{\rho} \bar{R}_{\rm in}^2 } 	\frac{g_e(x_e)}{g_e(x_{e0}) }		
%						- \Upsilon_{\rm ci} \frac{ \bar{T}^{5/2} }{ \bar{\rho} \bar{R}_{\rm in}^2 } 	\frac{g_i(x_i)}{g_e(x_{i0}) }
%						- \Upsilon_{\rm end} \frac{\bar{T}^{1/2} \bar{R}_c^2}{\bar{R}_{\rm in}^2} 
%						\right)		
%						\, \mathrm{d}\bar{t}
%					\right],
%	\label{eq:energy:EOS}
%\end{equation}
\begin{equation}
	\bar{U}
		=	\frac{\mathrm{CR}_{\rm in}^{4/3}}{\mathrm{CR}_{\rm in,preheat}^{4/3} }
			 \exp \left[ \int^{\bar{t}}_{\bar{t}_{\rm preheat}}  \left(
						 \Upsilon_\alpha 
						 			\, \bar{\rho} \, \bar{T}^{2.77}
						 			\eta_\alpha
						- \Upsilon_{\rm rad} \frac{\bar{\rho}}{\bar{T}^{1/2}} 
						{\color{black}- \Upsilon_{\rm c} \frac{ \bar{T} }{ \bar{B}_z \bar{R}_{\rm in}^2 } }
						- \Upsilon_{\rm end} \frac{\bar{T}^{1/2} \bar{R}_c^2}{\bar{R}_{\rm in}^2} 
						\right)		
						\, \mathrm{d}\bar{t}
					\right],
	\label{eq:energy:EOS}
\end{equation}
\end{widetext}
where $\mathrm{CR}_{\rm in,preheat}$ is the inner convergence ratio at the moment of preheat and is usually close to unity (to maximize pdV work done on the fuel).  In \Eq{eq:energy:EOS}, we also used the approximate power-law expression \eq{eq:basic:reactivity_simple} for the fusion reactivity.  The dimensionless parameters $\Upsilon_{\rm \alpha}$, $\Upsilon_{\rm rad}$, {\color{black}$\Upsilon_c$}, and $ \Upsilon_{\rm end}$ characterize the main energy source and loss mechanisms.  {\color{black}For hydrogen-like plasmas $(Z=1)$,} these parameters are
\begin{align}
	\Upsilon_{\alpha}	
			& \doteq	0.002
						\left[\rho_0 ({\rm g/cm^3})\right] \cdot
							 [T_{\rm preheat} (\mathrm{keV})]^{2.77}  \cdot [t_\varphi({\rm ns} )] ,
		\label{eq:energy:Upsilon_alpha} \\
	\Upsilon_{\rm rad}	
			& \doteq	{\color{black}0.67}~
						\frac{\left[\rho_0 ({\rm g/cm^3})\right] 
									\cdot [t_\varphi({\rm ns} )] }
						{{\color{black}A} \cdot [T_{\rm preheat} (\mathrm{keV})]^{1/2}  },
		\label{eq:energy:Upsilon_rad} \\
 	\Upsilon_c	
			& {\color{black}\doteq	0.04 ~
						\frac{ [T_{\rm preheat} (\mathrm{keV})] \cdot  [t_\varphi({\rm ns} )]}
						{ \left[ B_{z,0} ({\rm T})\right] 	\cdot [R_{\rm in,0}({\rm cm})]^2 },}
		\label{eq:energy:Upsilon_c}  \\
	\Upsilon_{\rm end} 
			& \doteq	{\color{black}0.05}~
						\frac{[T_{\rm preheat} ({\rm keV}) ]^{1/2}		
								\cdot  [t_\varphi({\rm ns} )]}
								{{\color{black}A^{1/2}} \cdot [h({\rm cm})] } .
		\label{eq:energy:Upsilon_end}	
\end{align}
%	\Upsilon_{\rm ce}	
%			& \doteq	5.3 \cdot  (10)^{-5} ~
%						\frac{ [T_{\rm preheat} (\mathrm{keV})]^{5/2} \cdot  [t_\varphi({\rm ns} )] \cdot g_e(x_{e0})   }
%						{ \ln \Lambda \cdot \left[ \rho_0 ({\rm g/cm^3})\right] 	\cdot [R_{\rm in,0}({\rm cm})]^2    } ,	
%		\label{eq:energy:Upsilon_ce} \\
% 	\Upsilon_{\rm ci}	
%			& \doteq	7.9 \cdot  (10)^{-7} ~
%						\frac{ [T_{\rm preheat} (\mathrm{keV})]^{5/2} \cdot  [t_\varphi({\rm ns} )]  \cdot  g_i(x_{i0})}
%						{ \ln \Lambda \cdot \left[ \rho_0 ({\rm g/cm^3})\right] 	\cdot [R_{\rm in,0}({\rm cm})]^2     },
%		\label{eq:energy:Upsilon_ci}  \\
{\color{black}As a side comment, when evaluating the relative importance of the energy gain and loss mechanisms at stagnation, the values that need to be substituted into \Eqs{eq:energy:Upsilon_alpha}--\eq{eq:energy:Upsilon_end} are the characteristic plasma parameters at stagnation, not the input parameters.  In particular, $t_\varphi$ is replaced by the burn-width time $\tau_{\rm bw}$.}

The dimensionless quantities \eq{eq:energy:Upsilon_alpha}--\eq{eq:energy:Upsilon_end} have different dependencies on the parameters describing the plasma fuel conditions ($R_{\rm in,0}$, $\rho_0$, $B_{z,0}$, $E_{\rm preheat}$, and $h$).  For example, $\Upsilon_\alpha$ and $\Upsilon_{\rm rad}$ are linear on the fuel density but have different nonlinear dependencies on the fuel temperature. 
%(In the high magnetization limit, $\Upsilon_{\rm ce}$ is also linear with plasma density.)  The parameters $\Upsilon_{\rm ce}$ and $\Upsilon_{\rm ci}$ have different power dependencies on the magnetic field $B_0$ when accounting for the magnetization regimes of the electrons and ions.  For example, near stagnation, $g_e \sim x_e^{-2}$ and $g_i \sim x_i^{-1}$, which leads to $\Upsilon_{\rm ce} \propto B_0^{-2}$ and $\Upsilon_{\rm ci} \propto B_0^{-1}$.  
{\color{black} The quantity $\Upsilon_{\rm c}$ is the only parameter depending on the magnetic field.}  Finally, only the end-loss parameter $\Upsilon_{\rm end}$ depends on the liner height $h$.  {\color{black} Since the dimensionless parameters $(\Upsilon_{\rm rad},\Upsilon_c,\Upsilon_{\rm end})$ have different dependencies, this will allow us to conserve the energy-loss mechanisms by carefully scaling the experimental input parameters for MagLIF.  This is discussed in \Sec{sec:scaling}.}

Equations \eq{eq:energy:Upsilon_alpha}--\eq{eq:energy:Upsilon_end} are helpful to evaluate the tradeoffs when making small changes to the input parameters of experimentally fielded MagLIF loads.  As an example, when substituting \Eq{eq:energy:T_preheat} into \Eqs{eq:energy:Upsilon_alpha}--\eq{eq:energy:Upsilon_end}, one finds that the parameters {\color{black}$\Upsilon_{\rm c}$} and $\Upsilon_{\rm end}$ increase with $\smash{T_{\rm preheat}}$ and therefore $\smash{\widehat{E}_{\rm preheat}}$.  Although increasing preheat will increase the initial adiabat of the fuel, this will not necessarily always lead to better performance for a MagLIF load since the relative loss rates due to {\color{black}thermal-conduction} and end flows increase.  Therefore, to make best use of gains in preheat energy delivery to MagLIF loads, one must also increase the magnetic field $B_{z,0}$ and the liner height $h$ so that {\color{black}thermal-conduction} losses and end-losses remain constant.

%%%%%%%%%%%%%%%%%%%%%%%%%%%%%%%%%%%%%%%%%%%%%%%%%
%%%%%%%%%%%%%%%%%%%%%%%%%%%%%%%%%%%%%%%%%%%%%%%%%
\section{Fuel mass losses}
\label{sec:mass}

MagLIF liners are open-ended to allow laser preheat of the fuel.  As the liner implodes, fuel mass can escape the imploding region of the liner.  The loss of fuel is modeled as a rarefraction wave.  Following \App{app:rarefraction}, we obtain
\begin{equation}
	\frac{\mathrm{d} \rho }{\mathrm{d}t} 
		\simeq	- 2 \rho \frac{1}{R_{\rm in}}\frac{ \mathrm{d} R_{\rm in}}{\mathrm{d}t}
				- 0.63  \frac{\rho c_s R_{\rm c}^2}{R_{\rm in}^2 h},
	\label{eq:mass:rho}
\end{equation}
which agrees with expressions reported in \Refs{McBride:2015ga,Slutz:2010hd} (modulo some minor numerical coefficients).  The first term in \Eq{eq:mass:rho} represents the adiabatic compression of the fuel.  The second term describes the fuel loss.  Equation~\eq{eq:mass:rho} can be integrated in order to get a generalized EOS for the fuel density:
\begin{equation}
	\frac{\rho(t) R_{\rm in}^2}{\rho_0 R_{\rm in,0}^2}
		=	\exp \left( -0.63 \int^t \frac{c_s}{h} \frac{R_{\rm c}^2}{R_{\rm in}^2} 
						\, \mathrm{d}t'
			\right),
	\label{eq:mass:EOS}
\end{equation}
where $\rho_0\doteq \rho(0)$ is the initial fuel mass density.  In the absence of fuel losses, the left-hand side of \Eq{eq:mass:EOS} denotes the increase of the fuel density with $\smash{\mathrm{CR}_{\rm in}}$ due to mass conservation.  When the liner has sufficiently imploded so that $R_{\rm c}=R_{\rm in}$, mass end losses are mainly determined by the ratio of the distance traveled by a characteristic sound wave ($\int^t c_s \, \mathrm{d} t)$ and the liner imploding height $h$.  Since  $c_s$ behaves approximately as $\smash{\mathrm{CR}_{\rm in}^{2/3}}$ to lowest order, the argument of the exponent is a relatively weak function of ${\rm CR_{\rm in}}$.  Hence, fuel-mass losses are distributed in time throughout the implosion.

In terms of dimensionless variables, \Eq{eq:mass:EOS} becomes
\begin{equation}
	\bar{\rho}
		=	\mathrm{CR}_{\rm in}^2
			\exp \left( - 0.70 \Upsilon_{\rm end}
							\int^{\bar{t}}
								  \frac{\bar{T}^{1/2} \bar{R}_{\rm c}^2}{\bar{R}_{\rm in}^2}
							\, \mathrm{d}\bar{t}'
			\right),
	\label{eq:mass:EOS_dim}
\end{equation}
where $\Upsilon_{\rm end}$ is given in \Eq{eq:energy:Upsilon_end}	.  Not surprisingly, the dimensionless parameter $\Upsilon_{\rm end}$ characterizes both energy and fuel--mass end losses {\color{black}since both processes are modeled as a rarefraction wave.}  In accordance to simple physical intuition, increasing the initial fuel temperature (for example, by increasing preheat) or making the liner implosions slower and longer in time will increase fuel mass losses.  To reduce relative end losses, the liner height $h$ must be increased.

%%%%%%%%%%%%%%%%%%%%%%%%%%%%%%%%%%%%%%%%%%%%%%%%%
%%%%%%%%%%%%%%%%%%%%%%%%%%%%%%%%%%%%%%%%%%%%%%%%%
\section{Magnetic field transport}
\label{sec:Bz}

{\color{black}
For the simple case of an axial magnetic field $\vec{B}=B_z(t,r)\vec{e}_z$ embedded in a cylindrical plasma column with radial velocity $\vec{v}=v_r(t,r)\vec{e}_r$ and temperature field $T=T(t,r)$, the induction equation for $B_z$ is
\begin{multline}
	\frac{\pd}{\pd t} B_z
		=	-\frac{1}{r}\frac{\pd}{\pd r} (r v_r B_z) 
			- \frac{1}{r} \frac{\pd}{\pd r} \left( r D_m \frac{\pd}{\pd r} B_z\right) \\
			- \frac{1}{e} \frac{1}{r} \frac{\pd}{\pd r} 
						\left( r \frac{\beta^{uT}_\wedge}{n_e} \frac{\pd}{\pd r} k_B T \right).
	\label{eq:Bz:induction}
\end{multline}
Here we assumed that axial gradients are negligible compared to radial gradients.  Due to the symmetry of the problem considered, the Hall term and Biermann-battery term are not considered.  The first term on the right-hand side of \Eq{eq:Bz:induction} is the advection term and is responsible for the compression of the magnetic flux as the liner implodes and for internal rearrangement of the magnetic-field profile within the liner cavity.  The second term is the magnetic-diffusion term which is responsible for the leakage of magnetic flux into the liner.  Here
\begin{equation}
	D_m \doteq \frac{m_e \nu_{ei}}{n_e e^2 \mu_0}
\end{equation}
is the magnetic diffusivity (assumed isotropic), $n_e = n_i \doteq \rho_i /m_i$ is the electron density, and $\nu_{ei}$ is the electron-ion collision frequency.  The last term in \Eq{eq:Bz:induction} is the Nernst-advection term where temperature gradients advect the magnetic field from the hot plasma core to the cold liner wall.  The parameter $\beta^{uT}_\wedge=\beta^{uT}_\wedge(x_e,n_e)$ is a magnetic-transport coefficient dependent on the electron Hall parameter and the electron density.  An expression for this coefficient is provided in Braginskii's transport model in Eq.~(4.31) of \Refa{Braginskii:1965vl}.

As discussed in \Refs{Vekshtein:1983aa,Vekshtein:1986aa,Velikovich:2015gs}, the magnetic flux is lost from the hot-fuel region at a faster rate compared to the loss rates simply estimated from the individual diffusion and Nernst terms in \Eq{eq:Bz:induction}.  Similarly to the enhanced thermal-conduction losses discussed in \Sec{sec:energy:sources_sinks}, as the fuel plasma is cooled by the liner wall, an internal convective flow arises within the isobaric plasma.  The internal flow is in the same direction as Nernst advection, so the magnetic field can effectively pile up near the liner wall where it forms a thin boundary layer.

For the problem of a semi-infinite plasma in contact with a cold liner wall, \Refa{Velikovich:2015gs} shows that the enhanced magnetic diffusion due to plasma flow and Nernst advection can be described using an effective magnetic diffusivity, which is given by
\begin{equation}
	D_{\rm m,eff} \doteq 8 D_B = \frac{k_B T}{2m_e\Omega_e}.
	\label{eq:Bz_Dmeff}
\end{equation}
Here $D_B$ is the Bohm-diffusion coefficient introduced in \Eq{eq:energy:DB}.  This effective magnetic diffusivity is found to be an accurate approximation for the case of magnetized ions in the plasma core so that $x_i>1$ and $x_e>30.3 A^{1/2}$ for hydrogen-like plasmas.

Based on the effective magnetic diffusivity \eq{eq:Bz_Dmeff}, we approximately describe the temporal dynamics of the magnetic field using the equation below:
\begin{equation}
	\frac{\pd}{\pd t} B_z
		=	-\frac{1}{r}\frac{\pd}{\pd r} (r v_r B_z) 
			- \frac{1}{r} \frac{\pd}{\pd r} \left( r D_{\rm m,eff} \frac{\pd}{\pd r} B_z\right) ,
	\label{eq:Bz:EOS_aux2}	
\end{equation}
where the magnetic advection term is kept in order to capture the compression of the magnetic flux as the liner implodes.  When spatially integrating \Eq{eq:Bz:EOS_aux2}, we obtain
\begin{equation}
	\frac{\mathrm{d} B_z}{\mathrm{d}t}
		\simeq	- 	2 \frac{B_z}{R_{\rm in}} \frac{ \mathrm{d} R_{\rm in}}{\mathrm{d}t}
				-	2 B_z \frac{D_{\rm m,eff}}{R_{\rm in}^2},
	\label{eq:Bz:EOS_aux3}
\end{equation}
where $v_r(R_{\rm in}) = \dot{R}_{\rm in}$ and we approximated $\pd_r B_z \simeq B_z/R_{\rm in}$.  In \Eq{eq:Bz:EOS_aux3}, $B_z=B_z(t)$ and $T=T(t)$ are considered spatially uniform.  Upon integrating \Eq{eq:Bz:EOS_aux3}, we obtain a generalized EOS for the axial magnetic field:
\begin{align}
	\frac{B_z(t) R_{\rm in}^2}{B_{z,0} R_{\rm in,0}^2}
		& =	\exp \left( - 2\int^t	 \frac{D_{\rm m,eff}}{R_{\rm in}^2} \, \mathrm{d}t' \right)	\notag \\
		& =	\exp \left( - \int^t	 \frac{k_B T}{e B_z R_{\rm in}^2} \, \mathrm{d}t' \right),
		\label{eq:Bz:EOS}
\end{align}
where we substituted \Eq{eq:Bz_Dmeff}.  In the absence of magnetic diffusion, we obtain $B_z(t) R_{\rm in}^2(t) = B_{z,0} R_{\rm in,0}^2$, which is the frozen-in law for ideal MHD.  As shown in \Eq{eq:Bz:EOS}, the effective magnetic diffusion leads to a loss in magnetic flux within the fuel and will therefore decrease the magnetic thermal insulation of the hot plasma core.  Interestingly, in the limit where magnetic-flux losses are small $(B_z R_{\rm in}^2 \simeq \const)$, the rate of magnetic-flux loss depends primarily on the fuel temperature.  For ideal adiabatic compression, the characteristic magnetic-flux confinement time $\tau_m$ depends on the inner convergence ratio as follows:
\begin{equation}
	\frac{1}{\tau_m} \doteq \frac{k_B T}{e B_z R_{\rm in}^2} \sim \mathrm{CR}_{\rm in}^{4/3}.
	\label{eq:tau_m}
\end{equation}
Hence, magnetic-flux losses are expected to increase as the liner implosion progresses.

In terms of dimensionless variables, \Eq{eq:Bz:EOS} is written as follows:
\begin{equation}
	\bar{B}
		=	\mathrm{CR}_{\rm in}^2 
				\exp \left( - \frac{\Upsilon_c}{4}
							\int^{\bar{t}} 
							\frac{\bar{T}}{ \bar{B}_z \bar{R}_{\rm in}^2}
							\, \mathrm{d}\bar{t}' 
				\right),
	\label{eq:Bz:EOS_dim}
\end{equation}
where $\Upsilon_c$ is the dimensionless parameter \eq{eq:energy:Upsilon_c} characterizing thermal-conduction losses.  It is not surprising that thermal and magnetic-flux losses are described by the same dimensionless parameter $\Upsilon_c$ since both of these loss mechanisms are enhanced by internal flows of the isobaric fuel column.\cite{Vekshtein:1983aa,Vekshtein:1986aa,Velikovich:2015gs}

The dimensionless quantity $\Upsilon_c/4$ can also be related to the \textit{inverse} of the magnetic Reynolds number Rm.  By making a change of variable $\mathrm{d}t = \mathrm{d} R_{\rm in} / \dot{R}_{\rm in}$ in \Eq{eq:Bz:EOS}, we may rewrite \Eq{eq:Bz:EOS_dim} as follows:
\begin{equation}
	\bar{B}
		=	\mathrm{CR}_{\rm in}^2 
				\exp \left( - \frac{1}{\rm Rm}
							\int^{\bar{R}_{\rm in}} 
							\frac{\bar{T}}{ \bar{B}_z \bar{R}_{\rm in}^2 \dot{\bar{R}}_{\rm in}}
							\, \mathrm{d}\bar{R}_{\rm in}
				\right),
	\label{eq:Bz:EOS_dim2}
\end{equation}
where the magnetic Reynolds number is
\begin{equation}
	\mathrm{Rm}	
		\doteq	0.01 \, \frac{ [B_{\rm z,0}R_{\rm in,0}({\rm T \, cm})] 
								 \cdot  [\dot{R}_{\rm in}({\rm km/s})] }
		 			{[T_{\rm preheat} ({\rm keV}) ]  }.
	\label{eq:Bz:Rm}
\end{equation}
In \Eq{eq:Bz:Rm}, $\dot{R}_{\rm in}$ should be interpreted as a characteristic implosion velocity.  Therefore, good confinement of the magnetic flux occurs when $\Upsilon_c\ll1$, or equivalently, $\mathrm{Rm}\gg 1$.  One can improve the confinement of the magnetic field by increasing the initial magnetic-field--radius product $\langle B_{\rm z,0}R_{\rm in,0}\rangle$, increasing the implosion velocity $\dot{R}_{\rm in}$, and decreasing the characteristic fuel temperature $T_{\rm preheat}$.}

%%%%%%%%%%%%%%%%%%%%%%%%%%%%%%%%%%%%%%%%%%%%%%%%%
%%%%%%%%%%%%%%%%%%%%%%%%%%%%%%%%%%%%%%%%%%%%%%%%%
\section{Fusion yield}
\label{sec:neutron}

For equimolar DT fuel, the neutron yield rate is approximately\cite{Atzeni:2009phys}
\begin{equation}
	\frac{\mathrm{d}Y}{\mathrm{d}t}
		=	\frac{1}{4} \left( \frac{\rho}{m_i} \right)^2 \, \langle \sigma v \rangle_{\rm DT} \, \pi R_{\rm in}^2 h,
	\label{eq:basic:yield_rate}
\end{equation}
where $\langle \sigma v \rangle_{\rm DT} (T)$ is the DT fusion reactivity. The Bosch--Hale expression for the fusion-reaction rate of DT plasma is\cite{Bosch:1992aa}
\begin{equation}
	\langle \sigma v \rangle_{\rm DT}
		= C_1 \frac{\xi^2}{\zeta^{5/6}} \exp\left( -3 \zeta^{1/3} \xi \right),
	\label{eq:basic:reactivity}
\end{equation}
where $\xi \doteq C_0 / T^{1/3}$ and 
\begin{equation}
	\zeta \doteq 1 - \frac{C_2 T + C_4 T^2 +C_6 T^3}{1+ C_3 T + C_5 T^2 +C_7 T^3}.
\end{equation}
The fit coefficients $C_i$ are given by $C_0 = 6.6610$ keV$^{1/3}$, $C_1=643.41 (10)^{-16}$ cm$^{3}$/s, $C_2= 15.136 (10)^{-3}$ keV$^{-1}$, $C_3 = 65.189(10)^{-3}$ keV$^{-1}$, $C_4 = 4.6054 (10)^{-3}$ keV$^{-2}$, $C_5 = 13.5 (10)^{-3}$ keV$^{-2}$, $C_6= -0.10675 (10)^{-3}$ keV$^{-3}$, and $C_7 = 0.01366 (10)^{-3}$ keV$^{-3}$.  Equation for the reactivity is valid within the $T=[0.2,100]$ keV range and has an error of less than 0.25\%.\cite{Atzeni:2009phys}  Within the 2--8 keV range, \Eq{eq:basic:reactivity} can be approximated by the following power law:
\begin{equation}
	\langle \sigma v \rangle_{\rm DT} \left( \frac{\rm cm^3}{s} \right)
		\simeq 2.89 \cdot (10)^{-20} \, [T({\rm keV})]^{3.77}.
	\label{eq:basic:reactivity_simple}
\end{equation}

Using \Eq{eq:basic:reactivity_simple}, we then write \Eq{eq:basic:yield_rate} in dimensionless form as
\begin{equation}
	\frac{\mathrm{d}Y}{\mathrm{d}\bar{t}}
		 =	Y_{\rm ref}  
		 			\bar{\rho}^2 \bar{T}^{3.77} \bar{R}_{\rm in}^2,
	\label{eq:dim:yield_rate}
\end{equation}
where the characteristic number for the yield is 
\begin{multline}
	Y_{\rm ref}
				 \doteq	1.30 \cdot  (10)^{18}
						\left[\rho_0 R_{\rm in,0}({\rm g/cm^2})\right]^2 	 		
						\cdot [ T ({\rm keV} )]^{3.77}		\\		
									\cdot [ h ({\rm cm} )] 
									\cdot [t_\varphi({\rm ns} )] .
	\label{eq:dim:Yref}
\end{multline}
This characteristic number will be used in Papers II and III to predict the scaling of the fusion yield.

%%%%%%%%%%%%%%%%%%%%%%%%%%%%%%%%%%%%%%%%%%%%%%%%%
\section{Summary of dimensionless parameters characterizing a MagLIF implosion}
\label{sec:summary}

\Tab{tab:parameters} summarizes the dimensionless parameters identified in the ODE-based model presented in Secs.~\ref{sec:electrical}--\ref{sec:neutron} describing a MagLIF implosion.  We now evaluate these parameters to compare their relative importance.

\begin{table*}
\caption{Dimensionless parameters characterizing the main physical process occurring in a MagLIF implosion.  The plasma conditions used to estimate some of the parameters below are shown in \Tab{tab:conditions}.} 
\label{tab:parameters}
%\begin{ruledtabular}
\begin{tabular}{ 
>{\centering\arraybackslash}m{0.15\linewidth} | 
>{\centering\arraybackslash} m{0.08\linewidth} |
>{\centering\arraybackslash} m{0.05\linewidth} |
m{0.4\linewidth}| 
>{\centering\arraybackslash} m{0.11\linewidth} |
>{\centering\arraybackslash} m{0.11\linewidth} 
}
\hline
\hline
Physics model  & Parameter   & Eq.  & \begin{center}
Description
\end{center}   & Characteristic value (preheat)  & Characteristic value (stagnation)  \\
\hline
Electrical circuit
						& $c_1$ 	&  \eq{eq:electrical:c1} 
 										&	Ratio of the outer-MITL inductance and total initial inductance of the pulsed-power generator 										
 										& \multicolumn{2}{c}{0.66} \\
 						& $c_2$   &  \eq{eq:electrical:c2}
 										&  Ratio of characteristic time $t_{\varphi}$ of the voltage source and the $(L_0+L_1)/Z_0$ time constant
 										& \multicolumn{2}{c}{1.23} \\
 						& $c_3$ 	&  \eq{eq:electrical:c3} 
 										&  Ratio of the $\sqrt{(L_0+L_1) C}$ time constant and the characteristic time $t_{\varphi}$ of the voltage source
 										& \multicolumn{2}{c}{0.01} \\
 										%& \checkmark  \\
 						& $c_4$ 	&  \eq{eq:electrical:c4} 
 						 				&  Measure of current transmission to the load region at early stages of the implosion
 										&  \multicolumn{2}{c}{549} \\
 						& $c_5$ 	&  \eq{eq:electrical:c5}
 						 				& Measure of current transmission to the load region at later stages of the implosion
 										& \multicolumn{2}{c}{1.71} \\
 						& $c_6$ 	&  \eq{eq:electrical:c6}
 						 				& Coupling parameter between imploding-load inductor and pulsed-power generator
 										& \multicolumn{2}{c}{0.14} \\
\hline
Liner implosion & $\Pi$ 	&  \eq{eq:liner:Pi}
 						 				&  Ratio of external magnetic potential energy and liner kinetic energy
 										& \multicolumn{2}{c}{5.55} \\
 						  & $\Phi$ &  \eq{eq:liner:Phi}
 						 				&  Ratio of fuel internal energy and liner kinetic energy
 										& \multicolumn{2}{c}{{\color{black}0.014}}\\
  						  & $\Lambda$ 
 						  				&  \eq{eq:liner:Lambda}
 						 				&  Ratio of liner internal energy and liner kinetic energy
 										& \multicolumn{2}{c}{0.614}\\
   						  & $\Sigma$ 
 						  				&  \eq{eq:liner:Sigma}
 						 				&  Ratio of internal magnetic potential energy and liner kinetic energy
 										& \multicolumn{2}{c}{$3.2 \cdot(10)^{-4}$} 	\\
\hline
Liner stability 		& $\Psi$ 	
 						  				&  \eq{eq:dim:Psi}
 						 				&  Measure of liner robustness towards instabilities
 										& \multicolumn{2}{c}{14.23} \\
\hline
Fuel energetics		& $\Upsilon_\alpha$,$\chi$ 	
										&  \eq{eq:energy:Upsilon_alpha}
 						 				&  Ratio of energy source due to $\alpha$ heating and fuel internal energy
 										& -
 										& {\color{black}0.07} 	\\
 						  & $\Upsilon_{\rm rad}$ 	
 						  				&  \eq{eq:energy:Upsilon_rad}
 						 				&  Ratio of radiation losses and fuel internal energy
 										& {\color{black}0.008}
 										& {\color{black}0.23} \\
 										%& \checkmark  \\
 						  & {\color{black}$\Upsilon_{\rm c}$ 	}
 						  				&  \eq{eq:energy:Upsilon_c}
 						 				&  Ratio of conduction losses and fuel internal energy
 										& {\color{black}0.02}
 										& {\color{black}1.08} \\
 						  & $\Upsilon_{\rm end}$ 	
 						  				&  \eq{eq:energy:Upsilon_end}
 						 				&  Ratio of end losses and fuel internal energy.  It is also a parameter for mass end losses.
 										&	{\color{black}0.06}
 										&  {\color{black}0.11}	 \\
 										%& \checkmark  \\
\hline
Particle transport	& $R_{\rm in}/\ell_\alpha$	
										&  \eq{eq:basic:a}
 						 				&  Ratio of plasma-column radius and stopping length of 3.5-MeV $\alpha$ particles
 										&  -
 										& {\color{black}0.05} 	\\
 						  & $R_{\rm in}/\varrho_\alpha$ 	
 						  				&  \eq{eq:basic:b}
 						 				&  Ratio of plasma-column radius and gyroradius of 3.5-MeV $\alpha$ particles
 										&   -
 										& {\color{black}1.48}\\
							& $x_\alpha$	
										&  \eq{eq:basic:x_alpha}
 						 				&  $\alpha$-trapping parameter
 										&  -
 										& {\color{black}0.32} 	\\
 						  & $\eta_\alpha$	
 						  				&  \eq{eq:basic:eta_alpha}
 						 				&  Fraction of trapped $\alpha$ particles
 										&   -
 										& {\color{black}0.27}\\
  						  & $x_e$ 	
 						  				&  \eq{eq:energy:xe}
 						 				& Electron magnetization
 										& {\color{black}0.28}
 										& {\color{black}58.0}	\\
\hline
{\color{black}Fuel mass losses }			& {\color{black}$0.7~\Upsilon_{\rm end}$}
						  				&  \eq{eq:energy:Upsilon_end}
 						 				&  {\color{black}Ratio of mass end losses and fuel mass}
 										&  {\color{black}0.04}
 										&  {\color{black}0.08}	 \\
\hline
Axial magnetic field transport
					 					&  {\color{black}$0.25~\Upsilon_c$}
 						  				&  \eq{eq:energy:Upsilon_c}
 						 				&  {\color{black}Ratio of effective magnetic diffusion and rate of magnetic flux compression}
 										&  {\color{black}0.005}
 										&  {\color{black}0.27}\\
\hline
Fusion yield 			& $Y_{\rm ref}$ 	
 						  				&  \eq{eq:dim:Yref}
 						 				&  Characteristic number for neutron yield
 										& 	-
 										& {\color{black}$1.35\cdot (10)^{15}$}	  \\
\hline
\hline
\end{tabular}
%\end{ruledtabular}
\end{table*}

For the circuit model, parameter values describing the Z pulsed-power generator are $Z_0=0.18~\Omega$, $L_0 = 9.58$~nH, $C = 0.1$~nF, and $L_1 = 5$~nH.  For the shunt resistor, typical values are $R_{\rm loss,i} = 80$~Ohm, $R_{\rm loss,f} = 0.25$~Ohm, $t_{\rm loss}=0$~ns, and $\Delta t_{\rm loss} = 5$~ns.\cite{foot:circuit}  As shown in \Fig{fig:electrical:Voc}, the characteristic timescale and amplitude of the voltage source are $t_{\rm \varphi} \simeq 100$~ns and $\varphi_0 \simeq 8.0$~MV, respectively.  MagLIF liners that are presently studied are $h=1$~cm long.  Upon substituting these values into \Eq{eq:electrical:Istar}, we obtain $I_\star \simeq 24.5$~MA for the characteristic current, which is close to the typical 20-MA peak current delivered to present-day MagLIF loads.\cite{Gomez:2020cd}  The evaluation of the dimensionless parameters $c_1$ to $c_6$ are given in \Tab{tab:parameters}.  $c_1\simeq 0.66$ indicates that the initial total inductance of the system mainly comes from the outer-MITL inductance $L_0$.  This is understandable since, as the load begins to implode, the inductance grows rapidly, so having $c_1 < 1$ is needed to have good dynamic-inductance matching and power delivery to the load when imploding at peak velocity.\cite{Katzenstein:1981jt} The parameter $c_2 \simeq 1.23$ indicates that the voltage drive of the Z circuit is close to resonant with the characteristic $(L_0+L_1)/Z_0$ time; therefore, the external voltage is driving in an efficient manner the ``oscillator" associated to the circuit.  The parameter $c_3 \simeq 0.01$ being small denotes that the role of the MITL capacitance in the circuit model in \Fig{fig:electrical:circuit} is minor.  This explains why including the MITL capacitance only adds small corrections to the calculation of the current delivered to the loads on the Z machine.\citep{McBride:2010hb}  The parameter $c_4\simeq 548$ being large means that current transmission is practically lossless in the early times of the current rise.  However, $c_5\simeq 1.71$ means that, at later times in the current pulse, the current transmission is lossy.  Finally, the parameter $c_6 \simeq 0.14$ indicates the pulsed-power generator is fairly stiff, and the level of coupling between the imploding load and the pulsed-power generator is relatively small (compared to lower-impedance university-scale generators).

Now, let us discuss the dimensionless parameters describing the liner implosion.  {\color{black}We consider the experimental input parameters for MagLIF shot z3289.\cite{Gomez:2020cd}}  The Be liner had radial dimensions $R_{\rm in,0} = 0.2325$~cm and $R_{\rm out,0} = 0.279$~cm so that $\mathrm{AR}=6$ and $\widehat{m}=0.138$~g/cm.  The preheat energy was {\color{black}$\smash{\widehat{E}_{\rm preheat} \simeq 1.2}$~kJ} and was delivered to a 1-cm long fuel cavity.  The initial fuel density  and external magnetic field were $\rho_0=1$~mg/cc and $B_{z,0}=16$~T, respectively.  {\color{black}Using the circuit parameters presented above, we find that the characteristic current is $I_\star \simeq 24.5$~MA and the characteristic timescale is $t_\varphi \simeq 100$~ns.  (The experimentally measured peak current  for z3289 was $\I\simeq$20~MA.\cite{Gomez:2020cd})}  For the Be EOS parameters, upon fitting a power law to a cold equation of state\cite{McBride:2015ga} for Be within the 1000-3000 Mbar range (typical stagnation pressures), we obtain  $p_{\rm ref} \doteq 66.91$~Mbar, $\rho_{\rm ref}\doteq 14$~g/cc, and $\gamma\doteq 1.923$. 
 
\Tab{tab:parameters} shows the evaluation of the dimensionless parameters $\Pi$, $\Phi$, $\Lambda$, and $\Sigma$ in \Eqs{eq:liner:Pi}--\eq{eq:liner:Sigma} when using these characteristic quantities.  $\Pi \simeq 5.55$ is the largest quantity since the magnetic drive is the largest energy source in the problem.  Surprisingly though, the second largest quantity is the parameter $\Lambda \simeq 0.614$, which is an order of magnitude larger than the parameter {\color{black}$\Phi \simeq 0.014$} characterizing the fuel internal energy.  This indicates that most of the magnetic energy in a MagLIF implosion is directed towards the compression of the liner, not to the fuel.  This observation agrees with results from numerical simulations.  For Be liners, the ratio $\Phi/\Lambda$ is
\begin{equation}
	\frac{\Phi}{\Lambda} 
		\simeq 0.003
			\frac{ [\widehat{E}_{\rm preheat} ({\rm kJ/cm})]  }{ [\widehat{m} ({\rm g/cm})] }.
\end{equation}
This ratio indicates that, in order to increase partitioning of energy towards the fuel, one must either increase the level of preheat per-unit-length $E_{\rm preheat}$ or reduce the mass of the imploding liner.  Regarding the latter option, decreasing the mass of the liners can lead to implosions that are less stable to MRTI.  Therefore, there is a tradeoff to be made.  Current work in underway to develop methods to mitigate instabilities, such as the use of coating on the external side of the liner, in order to access liner configurations that are lighter with higher initial aspect ratios.  Finally, $\Sigma \simeq 3.2\cdot(10)^{-4}$ denotes that most of the energy that compressed the MIF core region (fuel and magnetic field) is directed towards compressing the fuel.  This actually agrees with the general knowledge that MagLIF plasmas at stagnation generally tend to be high $\beta$ plasmas, \ie the magnetic pressure is small compared to the plasma pressure.

Let us now turn our attention to the evaluation of the dimensionless parameters characterizing the fuel energetics.  We evaluate these quantities by using the approximate plasma conditions given in \Tab{tab:conditions} for the preheat and stagnation phases of a MagLIF implosion.  We assume that the duration of the preheat stage is approximately {\color{black}7~ns}, and the duration of the stagnation stage is {\color{black}2.0~ns}.  Assuming that these conditions are also achievable in DT MagLIF experiments, we can evaluate the dimensionless parameters $\Upsilon_\alpha$, $\Upsilon_{\rm rad}$, {\color{black}$\Upsilon_{\rm c}$}, and $\Upsilon_{\rm end}$.\cite{foot:time_evaluation}  The results of these calculations are shown in \Tab{tab:parameters}.  

\begin{table}
\caption{Characteristic plasma conditions for the preheat and stagnation phases for MagLIF shot z3289.\cite{Gomez:2020cd}  Here $\Delta t$ represents a characteristic timescale of the process considered.  The preheat temperature is calculated from \Eq{eq:energy:T_preheat}.  {\color{black}The stagnation conditions were inferred using Bayesian data assimilation.\cite{foot:Knapp2022}  Note that the magnetic field at stagnation could not be measured from the diagnostic suite fielded for z3289, so the quoted value below is only approximate based on other similar MagLIF shots.}} 
\label{tab:conditions}
\begin{ruledtabular}
\begin{tabular*}{1 \textwidth}{ c c c c c c c c} 
Implosion 			&  $R_{\rm in}$  	&   $h$ 		&   $\rho$ 			&   $T$ 	&   $B_z$ & $\Delta t$ 	& $\ln \Lambda$ 	\\
phase 				&	(cm)					&	(cm)			&	(g/cm$^3$)		&	(keV)	&		(T)	& (ns)	\\
\hline
Preheat				& 0.2325							&	1						&	0.001		& {\color{black}0.06}					& 16 			& {\color{black}7}  	& 7 \\
Stagnation			& {\color{black}0.005}				&	1						&	{\color{black}0.7}		& {\color{black}2.7}					& {\color{black}$8\cdot 10^3$} 		&	{\color{black}2.0} & 7  \\
\end{tabular*}
\end{ruledtabular}
\end{table}

{\color{black}During the preheat stage, the main energy-loss mechanism is end losses, followed by thermal-conduction losses and radiation losses.  We note that thermal-conduction losses are likely being overestimated for the preheat stage since electrons are not highly magnetized and \Eq{eq:energy:chiperp} overestimates the thermal diffusivity in this regime.\cite{Velikovich:2015gs}  Estimating the thermal losses using the Braginskii thermal-transport coefficients\cite{Braginskii:1965vl} leads to $\sim10$x smaller values than what is reported in \Tab{tab:parameters} for the preheat stage.

At stagnation, we find that $\Upsilon_\alpha \simeq 0.07$ meaning low $\alpha$ heating for the DT plasma configuration in \Tab{tab:conditions}.  This is not surprising because the present-day Z facility does not have enough electrical energy and power to reach ignition-relevant conditions.  Regarding energy losses, the estimates in \Tab{tab:parameters} show that thermal conduction losses $(\Upsilon_c \simeq 1.08)$ are the most dominant in present-day MagLIF systems near stagnation.  Evaluation of radiation losses and end losses give $\Upsilon_{\rm rad} \simeq 0.23$ and $\Upsilon_{\rm end} \simeq 0.11$, respectively.  When scaling, it will be important to conserve the dimensionless parameters characterizing the energy-loss mechanisms so that the same physics regimes for energy transport are maintained.}

%radiation losses {\color{black}$(\Upsilon_{\rm rad} \simeq 0.227)$} are the most dominant in present-day MagLIF systems.  Therefore, when applying similarity scaling to MagLIF loads, one must ensure that the parameter $\Upsilon_{\rm rad}$ is conserved or does not become greater.\cite{Schmit:2020jd}  The parameter for ion-conduction losses is {\color{black}$\Upsilon_{\rm ci} \simeq 0.150$} and is much larger that that for electron-conduction losses.  This is not surprising since the high magnetic fields achieved via flux compression are sufficient to highly magnetize the electrons.  Finally, evaluation of the end losses shows that {\color{black}$\Upsilon_{\rm end} \simeq 0.106$, which is similar to $\Upsilon_{\rm ci}$.}  Based on these evaluations, we conclude that radiation losses, ion-conduction losses, and end losses are the main energy-loss mechanisms occurring near stagnation of a MagLIF implosion.
 
\Tab{tab:parameters} also shows the numerical evaluation of other dimensionless parameters characterizing particle-transport, magnetic-field transport, liner stability, and fusion yield.  For the parameters describing the confinement of $\alpha$ particles, we note that the $R_{\rm in}/\varrho_\alpha \gg R_{\rm in}/\ell_\alpha$ meaning that, if the same conditions where achieved in a MagLIF experiments using DT fuel, most of the confinement of the $\alpha$ particles is due to the flux-compressed magnetic field.  According to the parameter $\eta_\alpha$, roughly {\color{black}27\%} of $\alpha$ particles would be radially confined.  Regarding the electron Hall parameter, we observe that electrons are marginally magnetized during the preheat stage and become highly magnetized at stagnation.  {\color{black}Regarding magnetic-field transport, our estimates show that magnetic-flux losses are not negligible during near the stagnation phase of the implosion.  This is noteworthy since magnetic-flux losses degrade the magnetization of the hot plasma core and the confinement of thermal energy.}  Finally, substitution of the values given in \Tab{tab:conditions} into $Y_{\rm ref}$ {\color{black}gives good agreement with the observed DT-equivalent yields in present-day MagLIF experiments.}\cite{Gomez:2020cd}

%%%%%%%%%%%%%%%%%%%%%%%%%%%%%%%%%%%%%%%%%%%%%%%%%
%%%%%%%%%%%%%%%%%%%%%%%%%%%%%%%%%%%%%%%%%%%%%%%%%
%%%%%%%%%%%%%%%%%%%%%%%%%%%%%%%%%%%%%%%%%%%%%%%%%
\section{Similarity scaling MagLIF loads}
\label{sec:scaling}

%%%%%%%%%%%%%%%%%%%%%%%%%%%%%%%%%%%%%%%%%%%%%%%%%
\subsection{Definitions and assumptions}
\label{sec:scaling:philosophy}

Starting with some definitions, we say that two physical systems are \textit{dynamically similar} when the characteristics of one of them can be obtained from the characteristics associated with the other by means of a transformation of the input parameters of the problem that leaves the \textit{scale invariants} unchanged.\cite{Gratton:1991aa}  Scale invariants are always dimensionless quantities that are independent of the choice of the system of units.  In other words, two phenomena are similar if, and only if, all their dimensionless variables have the same numerical values.  With knowledge of the experimental inputs, the stagnation conditions, and performance of a ``baseline" MagLIF experiment, we can apply \textit{similarity scaling} to determine the input parameters of scaled MagLIF loads and provide analytical estimates of the stagnation conditions and performance.  Likewise, similarity scaling can be helpful to evaluate the veracity of MagLIF design simulations by comparing results corresponding to sets of carefully similarity-scaled MagLIF load designs.

To apply similarity scaling to MagLIF, we make the following first assumption:
\begin{enumerate}
\item We assume that the reduced models introduced in Secs.~\ref{sec:electrical}--\ref{sec:neutron} capture the main physical mechanisms at play in an imploding MagLIF load and are sufficient to qualitatively describe the \textit{trends} in performance metrics when varying the experimental parameters.  
\end{enumerate}
Of course, additional physical processes are missing: for example, skin effects in the current-driven liner, the finite time it takes for $\alpha$ particles to slow down, and radiation emission by mix.  We assume that the processes described in Secs.~\ref{sec:electrical}--\ref{sec:neutron} determine to leading order the dynamics of MagLIF implosions and that other missing physical processes provide next-order corrections.  Our assumption $\#1$ seems reasonable because the model proposed in this paper is a simplified version of the more sophisticated semi-analytical model in \Refa{McBride:2015ga}, which has been shown to sufficiently reproduce prediction trends of the more complex, multiphysics-code simulations (see \Refa{McBride:2016dm}).

One important consequence of Assumption $\#1$ is that any dimensionless quantity can be written as a relationship between scale invariants, \ie the identified dimensionless parameters.  More specifically, by inspection of the dimensionless equations presented in this paper, any dimensionless quantity $\bar{Q}$ can be determined by
\begin{multline}
	\bar{Q} = F_{\bar{\varphi}_{\rm oc},f} (c_{1-6}, \bar{t}_i, \Pi, \Phi,\Lambda, \Sigma, \Psi,
		\Upsilon_\alpha, 
		\Upsilon_{\rm rad}, {\color{black}\Upsilon_c},\Upsilon_{\rm end},...\\
		 R_{\rm in}/\ell_\alpha, R_{\rm in}/\varrho_\alpha, x_\alpha,\eta_\alpha),
	\label{eq:scaling:main}
\end{multline}
where the function $F_{\bar{\varphi}_{\rm oc},f}$ depends on the dimensionless time trace $\bar{\varphi}_{\rm oc}$ of the voltage drive and on the function $f$ parameterizing the behavior of the shunt resistor $R_{\rm loss}$.  ($\bar{t}_i$ denotes the dimensionless time parameters corresponding to $t_{\rm loss}$, $\Delta t_{\rm loss}$ and $t_{\rm preheat}$.)  As an example of how \Eq{eq:scaling:main} can be used, the dimensionless fuel pressure at stagnation is $\bar{p}_{\rm fuel,stag} \doteq \bar{p}_{\rm fuel}(\bar{t} =\bar{t}_{\rm stag})$.  {\color{black}From \Eqs{eq:liner:variables2} and \eq{eq:scaling:main},} the physical fuel pressure $p_{\rm fuel,stag}$ at stagnation will be given by
\begin{equation}
	\bar{p}_{\rm stag} 
		=	\frac{p_{\rm fuel,stag}}{p_{\rm preheat}}
		=	F_{\bar{\varphi}_{\rm oc},f} (c_{1-6}, \bar{t}_i, ... ,  x_\alpha,\eta_\alpha).
\end{equation}
Therefore, we have a relationship between the fuel pressure at stagnation, the pressure achieved during preheat [defined in \Eq{eq:liner:p_preheat}], and the function $F_{\bar{\varphi}_{\rm oc},f}$ of the scale invariants.  Note that $p_{\rm preheat}$ is function of the experimental input parameters so it is known \textit{a priori}.

We state our second assumption as follows:\\
\begin{enumerate}
\setcounter{enumi}{1}
\item No-$\alpha$ quantities, \ie quantities obtained in the absence of $\alpha$ heating, can be used as good surrogates for performance metrics of MagLIF configurations.
\end{enumerate}
For present-day MagLIF configurations fielded on the Z facility, the attained stagnation conditions are not sufficient for robust $\alpha$ heating.  Therefore, $\alpha$-heating can be neglected.  However, significant $\alpha$ heating and high yields could be achieved on a future larger pulsed-power driver.\cite{Slutz:2012gp,Sefkow:2014ik,McBride:2015ga,Slutz:2016cf,Slutz:2018iq}  For such higher-current scenarios, we shall use similarity scaling to estimate the stagnation conditions that are achievable in the absence of $\alpha$ heating and evaluate how promising they are to lead towards robust $\alpha$-heating scenarios.  Studying the scaling of no-$\alpha$ quantities was proposed in \Refa{Nora:2014iq} for laser-direct-drive ICF implosions.

We shall use the subscript ``no\,$\alpha$" to denote quantities in the absence of $\alpha$ heating.  The functional relationship for ``no\,$\alpha$" quantities is obtained by removing the dimensionless parameters related to $\alpha$ heating in \Eq{eq:scaling:main}:
\begin{multline}
	\bar{Q}_{\rm no\,\alpha} = \mc{F}_{\bar{\varphi}_{\rm oc},f} (c_{1-6},\bar{t}_i, \Pi, \Phi,\Lambda, \Sigma, \Psi,
		\Upsilon_{\rm rad}, {\color{black}\Upsilon_c}, \Upsilon_{\rm end}),
	\label{eq:scaling:main_noalpha}
\end{multline}
where $\mc{F}_{\bar{\varphi}_{\rm oc},f}$ denotes the function $F_{\bar{\varphi}_{\rm oc},f,{\rm no\,\alpha}}$.

Our third and final assumption is the following.
\begin{enumerate}
\setcounter{enumi}{2}
	\item Among the dimensionless ``no\,$\alpha$" parameters, the parameters that determine to leading order the dynamics and performance of a MagLIF implosion are: (i)~the parameters ($c_{1-6}$) determining the circuit-load coupling, (ii)~the $\Pi$ parameter determining the liner implosion, (iii)~the $\Phi$ parameter characterizing the preheat, (iv)~the parameter $\Psi$ measuring robustness towards hydrodynamic instabilities, and (v)~the parameters $(\Upsilon_{\rm rad},{\color{black}\Upsilon_c}, \Upsilon_{\rm end})$ characterizing the radiation, thermal-conduction, and end-loss mechanisms.
\end{enumerate}
It can be easily shown that, with the available experimental input parameters defining a MagLIF load, it is not possible to conserve all the dimensionless quantities on the right-hand side of \Eq{eq:scaling:main_noalpha}.  (Basically, the number of constraints to satisfy is larger than the number of independent experimental input parameters.)  To overcome this difficulty, we introduce the notion of \textit{essential dimensionless variables}.  A dimensionless parameter $\bar{Q}$ is considered essential when its value is neither too small or too large, \eg if its value lies within the range 0.1--10.  If the dimensionless parameter $\bar{Q}$ lies outside this range, it is assumed that its influence in the phenomenon is negligible, and it is therefore considered a \textit{non-essential dimensionless variable}.\cite{foot:nonessential}  {\color{black}(For an interesting discussion on essential and non-essential dimensionless variables, see \Refa{Gratton:1991aa}.)}  One example of a non-essential parameter is $\Sigma$, which measures the ratio of the axial magnetic energy within the fuel and the characteristic kinetic energy of the liner.  According to \Tab{tab:parameters}, $\Sigma$ is a small number for typical MagLIF implosions.  Therefore, conserving such parameter is not needed.  {\color{black}The parameter $c_3$ characterizing the effects of the MITL capacitance can be also considered a non-essential dimensionless parameter.  However, this is the sole parameter depending on the MITL capacitance $C$.  Therefore, when scaling, it is trivial to adjust $C$ in order to conserve $c_3$.}

%As shown in \Tab{tab:parameters}, ion-conduction losses dominate over electron-conduction losses at stagnation.  Since $\Upsilon_c$ and $\Upsilon_{\rm ci}$ are the sole energy-confinement parameters that depend on the applied magnetic field, we must then choose the scaling of $B_{z,0}$ so that $\Upsilon_{\rm ci}$ remains invariant.  

Although numerical evaluation of dimensionless parameters is a powerful tool to determine which dimensionless parameters are more essential than others, this procedure is sometimes insufficient.  In some cases, a MagLIF designer or experimentalist must tap into his/her physical intuition and decide which parameters are the most relevant to conserve when studying a physical mechanism at play.  When only a subset of the scale invariants appearing in a problem can be conserved simultaneously, we then have an \textit{incomplete (or partial) similarity scaling}.  {\color{black}When incompletely similarity scaling, there is a good chance that the scaling rules and subsequent scaling extrapolations will lead to predictable outcomes when the following two conditions are met.  First, the scaling rules conserve the essential scale invariants.  Second, the dimensionless parameters originally deemed non-essential do not need to be strictly conserved, but they must remain non-essential after the scaling is done.  When an originally non-essential parameter becomes essential, this can introduce non-accounted physics that can modify the predicted scaling rules for the metrics of the system.}  For MagLIF implosions, the dimensionless parameters mentioned in Assumption $\#3$ are those that (we assume) govern the main ingredients characterizing a MagLIF implosion: power delivery to the load, implosion and compression of the fusion fuel, and energy confinement of the assembled hot fuel.  Assumption $\#3$ will nevertheless need to be verified via numerical simulations of MagLIF loads and by experiments dedicated to studying scaling physics.

One mathematical consequence of Assumption $\#3$ is that the functional relation between dimensionless quantities can be simplified to
\begin{equation}
	\bar{Q}_{\rm no\,\alpha} \simeq 
		\mc{F}^{(0)}_{\bar{\varphi}_{\rm oc},f} (c_{1-6},\bar{t}_i, \Pi, \Phi,\Psi, 
		\Upsilon_{\rm rad}, {\color{black}\Upsilon_c},\Upsilon_{\rm end}),
	\label{eq:scaling:simple}
\end{equation}
where we replaced $\mc{F}_{\bar{\varphi}_{\rm oc},f}$ by the function $\mc{F}^{(0)}_{\bar{\varphi}_{\rm oc},f}$ which no longer has the parameters {\color{black}$\Sigma$ and $\Lambda$} in its arguments.  

There are several physical consequences of the approximation above.  Since the parameters $\Sigma$ and $\Lambda$ no longer appear, this implies that the potential energy of the flux-compressed magnetic field within the fuel and the internal energy of the liner are not {\color{black}considered when} determining conditions within the fuel.  Regarding the former, this is a good approximation since MagLIF plasmas are high-$\beta$ plasmas.  {\color{black}Regarding the latter, this is a much stronger and questionable approximation since the evaluation of the parameter $\Lambda$ in \Tab{tab:parameters} shows that $\Lambda$ is an essential dimensionless parameter and is not a negligible quantity.}  Therefore, dropping $\Lambda$ in \Eq{eq:scaling:simple} is not entirely warranted; this approximation will have to be verified \textit{a posteriori} in the scaling studies in Papers II and III.

After having mentioned our three assumptions, we are now in the position to discuss how similarity scaling will be used to explore the parameter space of MagLIF implosions.  Starting from the example given above, suppose that we are interested in determining the stagnation pressure of a scaled MagLIF load.  Based on the arguments above, the functional relationship for the stagnation pressure is approximately written as
\begin{equation}
	\frac{p_{\rm stag,no\,\alpha}}{p_{\rm preheat}}
		\simeq \mc{F}^{(0)}_{\bar{\varphi}_{\rm oc},f} (c_{1-6}, \bar{t}_i,\Pi, \Phi,\Psi, 
		\Upsilon_{\rm rad}, {\color{black}\Upsilon_c}, \Upsilon_{\rm end}).
	\label{eq:scaling:example}
\end{equation}

As shown in \Sec{sec:scaling:prescriptions}, the input parameters of the circuit and of a MagLIF load can be scaled so that the dimensionless quantities appearing on the right-hand side of \Eq{eq:scaling:example} are all conserved.  Therefore, the right-hand side of \Eq{eq:scaling:example} can be considered a constant so that
\begin{equation}
	p_{\rm stag,no\,\alpha}
		\simeq \frac{E_{\rm preheat}}{R_{\rm in,0}^2 h} \times \mathrm{const.} ,
\end{equation}
where we used \Eq{eq:liner:p_preheat}. This relationship holds true for any family of similarity-scaled MagLIF loads that are characterized by the same scale invariants $(c_{1-6}, \bar{t}_i,\Pi, \Phi,\Psi, \Upsilon_{\rm rad}, {\color{black}\Upsilon_c}, \Upsilon_{\rm end})$.  Suppose that the stagnation pressure $p_{\rm stag,no\,\alpha}$ is known for a ``baseline" MagLIF configuration.  If a scaled MagLIF configuration has the same scale invariants, then its stagnation pressure $p_{\rm stag,no\,\alpha}'$ will be given by
\begin{equation}
	p_{\rm stag,no\,\alpha}'
		=	p_{\rm stag,no\,\alpha}
			\left(\frac{E_{\rm preheat}'}{E_{\rm preheat}} \right)
			\left( \frac{R_{\rm in,0}^2 h}{R_{\rm in,0}'^2 h'} \right),
	\label{eq:scaling:pstag}
\end{equation}
where the quantities with apostrophes correspond to those of the scaled configuration.  In other words, by leveraging similarity scaling, we can estimate the solutions corresponding to a second set of input parameters by first taking the known baseline solution and multiplying it by a known function of the two sets of input parameters for different similarity-scaled configurations.  In other words, similarity scaling allows the projection of known solutions in a parameter space that has been explored to a new parameter space as long as the scale invariants describing the system remain unchanged.  In this example, we have taken the specific case of the stagnation pressure, but similar arguments hold for other experimental quantities of interest: for example, peak current, bang time, fuel temperature at peak burn, energy within the liner and the fuel, magnetic flux at stagnation, burn-width time, and neutron yield.  Applying similarity scaling therefore allows {\color{black}us} to reduce risks when doing extrapolations in performance by constraining the scaled loads to remain in a similar operating regime that is well understood experimentally or in numerical simulations.

%%%%%%%%%%%%%%%%%%%%%%%%%%%%%%%%%%%%%%%%%%%%%%%%%
\subsection{General similarity-scaling prescriptions}
\label{sec:scaling:prescriptions}

Let us now derive the scaling relations for the experimental input parameters characterizing the electrical circuit and a MagLIF load.  When enforcing conservation of the dimensionless parameters describing the circuit model [see \Eqs{eq:electrical:c1}--\eq{eq:electrical:c6}], we find
\begin{align}
	c_1\colon & \qquad \frac{L_1'}{L_1} 
			= \frac{L_0'}{L_0} 
			= \frac{h'}{h} ,
		\label{eq:scaling:L1} \\	
	c_2\colon & \qquad \frac{Z_0'}{Z_0} 
			= \frac{L_0'}{L_0}\frac{t_\varphi}{t_\varphi'}
			= \frac{h'}{h}\frac{t_\varphi}{t_\varphi'},	
		\label{eq:scaling:Z0} \\
	c_3\colon & \qquad \frac{C'}{C} 
			= \frac{L_0}{L_0'}  \left(\frac{t_\varphi'}{t_\varphi}\right)^2
			= \frac{h}{h'} \left(\frac{t_\varphi'}{t_\varphi}\right)^2,
		\label{eq:scaling:C} \\
	c_4\colon & \qquad \frac{R_{\rm loss,i}'}{R_{\rm loss,i}} 
			= \frac{L_0'}{L_0}\frac{t_\varphi}{t_\varphi'} 
			= \frac{h'}{h}\frac{t_\varphi}{t_\varphi'} ,	
		\label{eq:scaling:Rloss_i} \\			
	c_5\colon & \qquad \frac{R_{\rm loss,f}'}{R_{\rm loss,f}} 
			= \frac{L_0'}{L_0}\frac{t_\varphi}{t_\varphi'} 
			= \frac{h'}{h}\frac{t_\varphi}{t_\varphi'} ,	
		\label{eq:scaling:Rloss_f} \\						
	c_6\colon & \qquad \frac{L_0'}{L_0} 
			= \frac{h'}{h} .
		\label{eq:scaling:L0} 
\end{align}
All temporal variables should scale with the voltage characteristic time.  Therefore, the time parameters describing the shunt resistor are scaled as
\begin{equation}
	\bar{t}_i \colon \qquad
	 \frac{t_{\rm loss}'}{t_{\rm loss}}
		=	\frac{t_\varphi'}{t_\varphi},
	\qquad
	\frac{\Delta t_{\rm loss}'}{\Delta t_{\rm loss}}
		=	\frac{t_\varphi'}{t_\varphi}.
	\label{eq:scaling:tloss} 
\end{equation}
As a reminder, for an arbitrary quantity $Q$ corresponding to the \emph{baseline} set of input parameters, the quantity $Q'$ denotes its \textit{scaled} value for the second set of input parameters.  

The origin of the scaling prescriptions in \Eqs{eq:scaling:L1}--\eq{eq:scaling:L0} can be understood from the effects that the parameters $c_1$ to $c_6$ measure.  As an example, in Papers II and III, we investigate similarity scaling of MagLIF loads when the peak current and characteristic timescale $t_\varphi$ of the external voltage source are varied.  As shown in those papers, the target height changes when scaling according to these parameters.  To maintain the relative impedance of the electrical circuit when imploding shorter or longer targets, the dimensionless quantity $c_6$ constrains that the circuit inductance $L_0$ scale with the target height.  The rest of the scaling prescriptions in \Eqs{eq:scaling:L1}--\eq{eq:scaling:tloss} follow from similar arguments.

In scaling studies of MIF concepts to larger and more energetic pulsed-power drivers, the extrapolation is usually done in terms of the peak current delivered to the load (e.g., see \Refs{Bures:2012ia,Velikovich:2007hq}). However, peak current delivered to the load is a result from the solution of the electrical circuit equations, and therefore, does not suit well as an independent variable for similarity scaling.  Luckily, the characteristic current $I_\star$ in \Eq{eq:electrical:Istar} can serve as such variable.  Noting that $I_\star = (\varphi_0 t_\varphi / L_0) [c_1/(1+c_2)]$ and using conservation of the parameters $c_1$ and $c_2$, we find the scaling prescription for the characteristic voltage $\varphi_0$:
\begin{equation}
	\frac{\varphi_0'}{\varphi_0}
		=	\frac{I_\star'}{I_\star} \frac{t_\varphi}{t_\varphi'}  \frac{L_0'}{L_0}
		=	\frac{I_\star'}{I_\star} \frac{t_\varphi}{t_\varphi'}  \frac{h'}{h}. 
\end{equation}
All measured voltages in the circuit model of \Fig{fig:electrical:circuit} will scale according to the prescription above.  As expected, the voltage $\varphi_0$ increases when the characteristic current increases.  It also increases when the characteristic time $t_\varphi$ decreases because of the larger inductive voltage.  Finally, when driving shorter or longer liners, the voltage $\varphi_0$ will also scale proportionally to $h$.  

Among the dimensionless parameters describing the liner dynamics, the most important quantity to conserve is the $\Pi$ parameter, which characterizes the magnetic drive of the implosion.  We also wish to conserve the robustness of the MagLIF liner towards MRTI when scaling the input parameters.  Thus, we have $\Pi = \const$ and $\Psi = \const$.  When combining these two constraints, we find that the liner outer radius and the liner mass per-unit-length must scale as follows:
\begin{align}
	\Pi~\&~\Psi\colon & \quad\frac{R_{\rm out,0}'}{R_{\rm out,0}}
			=	\bigg[
					\left( \frac{I_\star'}{I_\star} \right)^{1-1/\gamma}
					\left( \frac{t_\varphi'}{t_\varphi} \right)
				\bigg]^{\frac{1}{2-1/\gamma}}  ,
		\label{eq:scaling:Rout}  \\
	\Pi~\&~\Psi\colon & \qquad\frac{\widehat{m}'}{\widehat{m}}
			=	\bigg[
					\left( \frac{I_\star'}{I_\star} \right)^2
					\left( \frac{t_\varphi'}{t_\varphi} \right)^{2-2/\gamma}
				\bigg]^{\frac{1}{2-1/\gamma}} .
		\label{eq:scaling:mhat} 
\end{align}
Following these scaling rules guarantees that the liner will implode in a similar fashion and that its robustness towards instabilities will be maintained.  For the sake of completeness, the liner inner radius will then follow
\begin{equation}
	R_{\rm in,0}' = \left[ R_{\rm out,0}'^2 - \widehat{m}'/(\pi \rho_{\rm liner,0}) \right]^{1/2}.
	\label{eq:scaling:Rin}
\end{equation}

By conservation of the parameter $\Phi$ in \Eq{eq:liner:Phi}, which determines the pushback of the fuel pressure on the imploding liner, we find that the scaling relation for the preheat energy per-unit-length is given by
\begin{equation}
	\Phi\colon \qquad
		\frac{\widehat{E}_{\rm preheat}'}{\widehat{E}_{\rm preheat} }
		=	 \left( \frac{I_\star'}{I_\star} \right)^2.
	\label{eq:scaling:Epreheat_hat}
\end{equation}
This relation can be found by dividing the two constant parameters $\Pi$ and $\Phi$.  We therefore find that the preheat energy per-unit-length scales as the characteristic current squared.  Since all time parameters scale with the characteristic timescale of the voltage drive, then the preheat timing will also scale accordingly:
\begin{equation}
	\bar{t}_i \colon \qquad
		\frac{t_{\rm preheat}'}{t_{\rm preheat}}
			=	\frac{t_\varphi'}{t_\varphi}.
\end{equation}
Physically, this scaling law ensures that preheat will occur at the same moment of the implosion (in terms of the inner convergence ratio $\smash{\mathrm{CR}_{\rm in}}$).

We still need to determine the scaling prescriptions for the initial fuel density $\rho_0$, the applied external magnetic field $B_{z,0}$, and the liner height $h$.  From the discussion in \Sec{sec:scaling:philosophy}, the scaling prescriptions for these input parameters are determined by fixing the dimensionless parameters describing relative radiation losses, thermal-conduction losses, and end losses.  We then obtain the following scaling rules:
\begin{align}
	\Upsilon_{\rm rad}\colon & \quad
	\frac{\rho_0'}{\rho_0}
			=	\left( \frac{I_\star'}{I_\star} \right)^{2/3}
				\left( \frac{R_{\rm in,0}'}{R_{\rm in,0} } \right)^{-2/3}
				\left( \frac{t_\varphi'}{t_\varphi } \right)^{-2/3}   ,
		\label{eq:scaling:rho}  \\
	{\color{black}\Upsilon_c}\colon  & \quad 
	\frac{B_{z,0}'}{B_{z,0}} 
			=	\left( \frac{I_\star'}{I_\star} \right)^{4/3}
				\left( \frac{R_{\rm in,0}'}{R_{\rm in,0} } \right)^{-10/3}
				\left( \frac{t_\varphi'}{t_\varphi } \right)^{5/3},
		\label{eq:scaling:Bz} \\
	\Upsilon_{\rm end}\colon & \quad 
	\frac{h'}{h}
			=	\left( \frac{I_\star'}{I_\star} \right)^{2/3}
				\left( \frac{R_{\rm in,0}'}{R_{\rm in,0} } \right)^{-2/3}
				\left( \frac{t_\varphi'}{t_\varphi } \right)^{4/3}.
		\label{eq:scaling:h} 
\end{align}
To obtain these scaling prescriptions, we have used \Eq{eq:energy:T_preheat} to eliminate the preheat temperature $T_{\rm preheat}$ in favor of the preheat energy per-unit-length $\smash{\widehat{E}_{\rm preheat}}$.  We then substituted \Eq{eq:scaling:Epreheat_hat}.  Finally, the scaling of the preheat energy delivered to the MagLIF fuel volume is given by
\begin{equation}
	\frac{E_{\rm preheat}'}{E_{\rm preheat}}
		=	\frac{\widehat{E}_{\rm preheat}'}{\widehat{E}_{\rm preheat}}
			\frac{h'}{h}.
	\label{eq:scaling:Epreheat}
\end{equation}

In summary, \Eqs{eq:scaling:L1}--\eq{eq:scaling:Epreheat} are the scaling prescriptions for MagLIF experimental input parameters when varying the characteristic current (or equivalently, the peak current) delivered to the load and when varying the characteristic time of the voltage drive (or equivalently, the rise time of the current trace).  In Papers II and III, we shall use these scaling prescriptions and test similarity scaling according to $I_\star$ and $t_\varphi$ via numerical simulations.  

{\color{black}
%%%%%%%%%%%%%%%%%%%%%%%%%%%%%%%%%%%%%%%%%%%%%%%%%
\subsection{Example}
\label{sec:scaling:example}

In this section, we provide a brief example on how the scaling theory can be applied to MagLIF.  More details on the example below are given in Paper II.  We consider scaling a MagLIF load with respect to the characteristic current $I_\star$ while maintaining the implosion timescale fixed $(t_\varphi=\const)$.  To determine the scaling rules for the MagLIF input parameters, we need to first determine the baseline liner configuration.  We consider a Be liner with initial density 1.858~g/cm$^3$ and initial radial dimensions $R_{\rm in,0}=2.325$~mm and $R_{\rm out,0}=2.79$~mm.  This corresponds to an initial AR=6 and mass per-unit-length $\widehat{m}$=139~mg/cm.  The baseline liner is driven at 20~MA, and the polytropic index is chosen to be $\gamma=2.25$.  When including a small correction to \Eq{eq:scaling:Rout} to account for differences in the initial shock compression of the liner (see Paper II), we obtain the following scaling laws for the liner inner and outer radii:
\begin{equation}
	\frac{R_{\rm out,0}'}{R_{\rm out,0}} \simeq \left( \frac{I_\star'}{I_\star} \right)^{0.381}, 	\qquad
	\frac{R_{\rm in,0}'}{R_{\rm in,0}} \simeq \left( \frac{I_\star'}{I_\star} \right)^{0.206}.  
	\label{eq:scaling:example_R}
\end{equation}
To simplify the upcoming calculation, here we fitted power laws to the exact scaling rules in \Eqs{eq:scaling:Rout}--\eq{eq:scaling:Rin}.  Upon substituting \Eqs{eq:scaling:example_R} into \Eqs{eq:scaling:rho}--\eq{eq:scaling:h} and \eq{eq:scaling:Epreheat}, we obtain the following power-law scaling laws for the preheat energy $E_{\rm preheat}$, initial fuel density $\rho_0$, magnetic field $B_{z,0}$, and liner height $h$:
\begin{gather}
	\frac{E_{\rm preheat}'}{E_{\rm preheat}} \simeq \left( \frac{\I'}{\I} \right)^{2.529} , 
		\label{eq:scaling:example_Epreheathat}\\
	\frac{\rho_0'}{\rho_0} = \frac{h'}{h}  \simeq \left( \frac{\I'}{\I} \right)^{0.529}, 
		\label{eq:scaling:example_rho}\\
	\frac{B_{z,0}'}{B_{z,0}}   \simeq \left( \frac{\I'}{\I} \right)^{0.647} .
		\label{eq:scaling:example_Bz}
\end{gather}

From the scaling laws \eq{eq:scaling:example_R}--\eq{eq:scaling:example_Bz}, we may now determine how the no-$\alpha$ stagnation pressure $p_{\rm stag,no\,\alpha}$ scales for this baseline liner configuration when increasing the characteristic current.  Substituting \Eqs{eq:scaling:example_R}--\eq{eq:scaling:example_Bz} into \Eq{eq:scaling:pstag} leads to 
\begin{equation}
	\frac{p_{\rm stag,no\,\alpha}'}{p_{\rm stag,no\,\alpha}}
		\simeq
			\left(\frac{E_{\rm preheat}'}{E_{\rm preheat}} \right)
			\left( \frac{R_{\rm in,0}^2 h}{R_{\rm in,0}'^2 h'} \right)
		\simeq \left( \frac{I_\star'}{I_\star} \right)^{1.59}.
	\label{eq:scaling:example_pstag}
\end{equation}
Hence, the no-$\alpha$ stagnation pressure is expected to increase by a factor of $(60/20)^{1.59}\simeq5.7$ when similarity scaling from 20~MA to 60~MA drive current.

Figure~\ref{fig:pion} compares the analytical scaling law \eq{eq:scaling:example_pstag} to the simulated plasma pressure for a family of similarity-scaled MagLIF configurations.  The simulation results were obtained from 2D ``clean" calculations using the radiation-magnetohydrodynamics code \textsc{hydra}.\cite{Marinak:1996fs,Koning:2009}  As shown, the analytical scaling curve \eq{eq:scaling:example_pstag} agrees with the simulation results when scaling from 15~MA to 60~MA peak current.  This is a noteworthy because \Eq{eq:scaling:example_pstag} only depends on the input parameters while the stagnation pressure is a ``end result" of several physical processes occurring during a MagLIF implosion.  This agreement is achieved by conserving the dimensionless parameters characterizing the system.  The scaling theory does not include $\alpha$ heating.  Thus, it is not surprising that the simulation results with $\alpha$ heating eventually deviate from the no-$\alpha$ analytical scaling rule when $\alpha$ heating becomes relevant.  According to these calculations, this occurs when driving a MagLIF load above 40~MA.

Paper II of this series provides a more thorough study on the similarity-scaling of MagLIF loads with respect to peak current.  In Paper II, the implosion kinematics of similarity-scaled MagLIF configurations are compared, the scaling of other stagnation metrics (\eg, ion temperature and $R_{\rm in}/\varrho_\alpha$) are discussed, and the scaling of the performance metrics such as the fusion yield and the Lawson ignition parameter $\chi$ are shown.\cite{Betti:2010fc}
}

\begin{figure}
	\includegraphics[scale=.43]{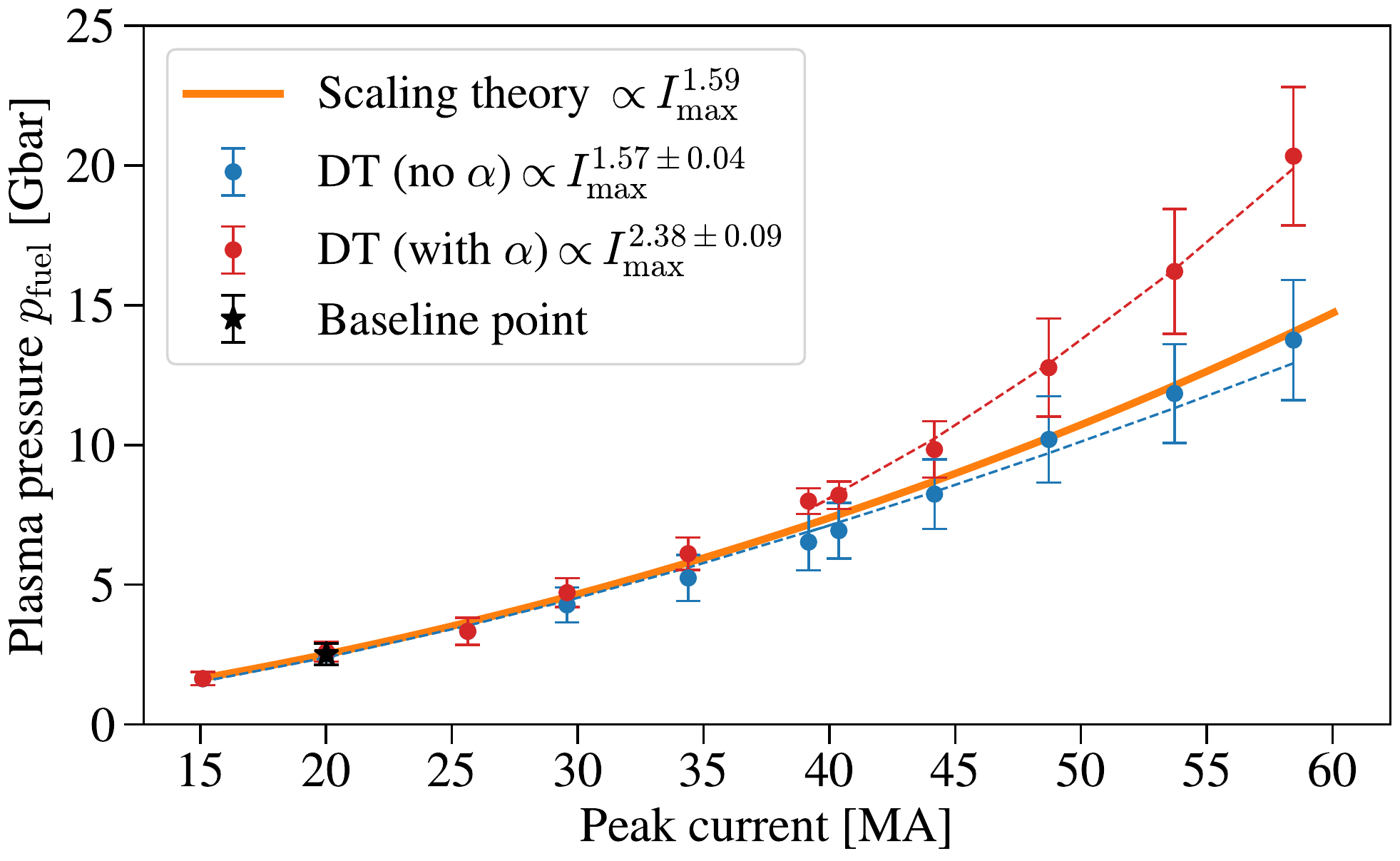}
	\caption{{\color{black}Burn-history averaged plasma pressure corresponding to a family of similarity-scaled MagLIF configurations.  The red and blue points denote simulation results with and without $\alpha$ heating, respectively.  Dashed lines are power-law fits to the simulation data.  The legend shows the fitted scaling exponents.  Error bars denote the burn-weighted standard deviation associated to temporal variations of the plasma pressure near peak burn.  The orange curve is the theoretical scaling law in \Eq{eq:scaling:example_pstag}. } }
	\label{fig:pion}
\end{figure}

%%%%%%%%%%%%%%%%%%%%%%%%%%%%%%%%%%%%%%%%%%%%%%%%%
%%%%%%%%%%%%%%%%%%%%%%%%%%%%%%%%%%%%%%%%%%%%%%%%%
%%%%%%%%%%%%%%%%%%%%%%%%%%%%%%%%%%%%%%%%%%%%%%%%%
\section{Conclusions}
\label{sec:conclusions}

Pulsed-power driven MagLIF loads are complex systems that involve multiple processes that govern the electrical-circuit dynamics, liner-implosion dynamics, hydrodynamic instabilities, fuel-energy losses, particle transport, fuel-mass losses, axial magnetic-field transport, fusion yield, etc.  Moreover, the experimental input parameters defining a MagLIF load are numerous.  Given the complexity of MagLIF implosions and the relatively high dimensionality of the input-parameter space, detailed explorations of MagLIF performance is computationally expensive using radiation-hydrodynamic simulations and virtually impossible via integrated experiments on Z. However, surveys of MagLIF performance in the input-parameter space can be simplified via the use of \textit{dimensional analysis} and \textit{similarity scaling}.

In Paper I of this series, we presented a theoretical framework for similarity scaling MagLIF loads.  We developed a simplified analytical model based on ordinary differential equations (ODEs) that describes the main mechanisms at play occurring during a MagLIF implosion.  After introducing the governing equations, we performed a dimensional analysis to determine the most important dimensionless parameters characterizing MagLIF implosions.  We also provided estimates of such parameters using typical fielded or experimentally observed stagnation quantities for MagLIF.  We then showed that MagLIF loads can be \textit{incompletely similarity scaled}, meaning that the experimental input parameters of MagLIF can be varied such that many (but not all) of the dimensionless quantities characterizing a MagLIF implosion are conserved.  Following arguments based on similarity scaling, we can then estimate the performance of similarity scaled MagLIF loads when using this framework. 

Specific applications of these results are left to \Refs{foot:Ruiz_current,foot:Ruiz_time} (also referred as Papers II and III of this series).  In Paper~II, we address the problem of scaling MagLIF targets with respect to electrical current delivered to the load.  In Paper~III, we study the scaling of MagLIF loads when the characteristic timescale of the voltage drive (or equivalently, the current rise-time) is varied.   
%Finally, in paper IV,\cite{foot:Ruiz_mR2} we address the scaling of a MagLIF liner radius and mass in order to increase the robustness of liners towards magneto-Rayleigh--Taylor (MRT) instabilities. 

The present work can be extended in several directions.  First, there are ``hidden" dimensionless variables that are not captured in the present model.  Examples of known physical processes that are not discussed are: (i)~magnetic diffusion of the external magnetic field into the liner, (ii)~magnetic-field advection due to Hall physics related to the tenuous plasmas surrounding the MagLIF liner, (iii)~the role of the {\color{black}three-dimensional} helical structures observed in MagLIF-like implosions {\color{black}which can lead to complex velocity-shear flow},\cite{Awe:2014gba,Awe:2013dt} (iv)~the electro-thermal instability,\cite{Oreshkin:2008dy,Peterson:2012bu,Awe:2021jp,Peterson:2013eh,Yu:2020gl} (v)~{\color{black}the introduction of mix by the interaction of the preheat laser with the laser-entrance-hole (LEH) window and the cushions,\cite{Knapp:2019gf} the impact of the preheat blast wave with the inner liner wall, the ablation of liner material,\cite{Garcia-Rubio:2017aa,Garcia-Rubio:2018aa}, and deceleration-type instabilities near stagnation,\cite{Knapp:2017jw}} and (vi)~degradation of $\alpha$-particle confinement due to liner instabilities.  Understanding how these processes affect MagLIF performance and how they scale remains an important task.  Second, future scaling studies can be more focused towards the physics of the laser-preheat process itself.  In \Refa{Schmit:2020jd}, a preliminary scaling analysis was presented on the estimated stopping length of the laser and laser--plasma interactions.  This analysis could be further improved and corroborated against numerical simulations, as well as dedicated experiments.\cite{foot:Pollock}  Finally, dimensional analysis and similarity scaling may be useful for understanding the physics behind the pulsed-power delivery to a MagLIF load.  This effort would involve developing a suitable analytical model for describing current losses in the magnetically-insulated transmission lines.  Such a study could then provide guidance on risks associated to pulsed-power delivery when extrapolating to higher current densities in a future pulsed-power machine.

One of the authors (D.~E.~Ruiz) was supported in part by Sandia National Laboratories (SNL) Laboratory Directed Research and Development (LDRD) Program, Project 223312.  Sandia National Laboratories is a multimission laboratory managed and operated by National Technology $\&$ Engineering Solutions of Sandia, LLC, a wholly owned subsidiary of Honeywell International Inc., for the U.S. Department of Energy's National Nuclear Security Administration under contract DE-NA0003525.  This paper describes objective technical results and analysis. Any subjective views or opinions that might be expressed in the paper do not necessarily represent the views of the U.S. Department of Energy or the United States Government.

%%%%%%%%%%%%%%%%%%%%%%%%%%%%%%%%%%%%%%%%%%%%%%%%%
%%%%%%%%%%%%%%%%%%%%%%%%%%%%%%%%%%%%%%%%%%%%%%%%%
%%%%%%%%%%%%%%%%%%%%%%%%%%%%%%%%%%%%%%%%%%%%%%%%%

\appendix

%%%%%%%%%%%%%%%%%%%%%%%%%%%%%%%%%%%%%%%%%%%%%%%%%
\section{Rarefraction wave}
\label{app:rarefraction}

We model the flow outflow from the openings of the imploding liner as a rarefraction wave.  The 1D Euler equations are
\begin{gather}
	\pd_t \rho + u\pd_x  \rho = - \rho \pd_x u,\\
	\rho \pd_t u + \rho u \pd_x  u = - \pd_x p,\\
	\pd_t S + u \pd_x  S = 0,
\end{gather}
where $\rho(t,x)$, $u(t,x)$, $p(t,x)$, and $S(t,x)$ are the fluid density, velocity, pressure, and entropy, respectively.  Following \Refa{foot:Zeldovich}, we propose self-similar solutions of the form $f=f(\xi)$, where $\xi \doteq x/t$.  In terms of $\xi$, the 1D Euler equations for isentropic flow are
\begin{gather}
	(u-\xi) \frac{\mathrm{d} \rho}{\mathrm{d}\xi} = - \rho \frac{\mathrm{d} u}{\mathrm{d} \xi} ,
		\label{eq:app:rho_trans}	\\
	\rho (u-\xi) \frac{\mathrm{d} u}{\mathrm{d}\xi}= - \frac{\mathrm{d} p}{\mathrm{d} \xi},
		\label{eq:app:u_trans}\\
	(u-\xi) \frac{\mathrm{d} S}{\mathrm{d}\xi} = 0.
\end{gather}
The last equation implies that $\mathrm{d}S/\mathrm{d}\xi=0$.  Combining the first and second equations, we obtain
\begin{equation}
	(u-\xi)^2 \frac{\mathrm{d} \rho}{\mathrm{d}\xi} 
		=	 \frac{\mathrm{d} p}{\mathrm{d} \xi}
		= 	 \left( \frac{\mathrm{d} p}{\mathrm{d} \rho} \right)_S \frac{\mathrm{d} \rho}{\mathrm{d}\xi} 
		=  c^2 \frac{\mathrm{d} \rho}{\mathrm{d}\xi} .
	\label{eq:app:main_eq}
\end{equation}
Because the flow is isentropic, the pressure $p$ is only dependent on the fluid density: $p=p(\rho)$, and the local sound speed is denoted by $\smash{c\doteq ( \pd p/\pd \rho)_S^{1/2}}$.  A nontrivial solution for \Eq{eq:app:main_eq} is given by
\begin{equation}
	u - \xi = \pm c.
	\label{eq:app:u_aux}
\end{equation}
Substituting this result into \Eqs{eq:app:rho_trans} and \eq{eq:app:u_trans} gives
\begin{equation}
	\mathrm{d}u \pm c \frac{\mathrm{d} \rho}{\rho } 
		=	\mathrm{d}u \pm \frac{\mathrm{d} p}{\rho c} 
		=	0.
	\label{eq:app:invariants_aux}
\end{equation}
Therefore, the following two functions are constants:
\begin{equation}
	J_\pm 	\doteq u \pm \int  \frac{\mathrm{d} p}{\rho c}  
				= u \pm \int c \frac{\mathrm{d} \rho}{\rho } = \const.
	\label{eq:app:J}
\end{equation}
For an ideal gas, the integral quantities $\int \mathrm{d}p/(\rho c) = \int c \, \mathrm{d}\rho / \rho$, can be expressed in terms of the local speed of sound $c$ alone.  For example, for a perfect gas with constant specific heats and $\gamma=5/3$, we have
\begin{align}
	p & = p_0 \left( \frac{\rho}{\rho_0}\right) ^{5/3}, \\
	c^2 & = \frac{\pd p}{\pd \rho } = c_0^2 \left( \frac{\rho}{\rho_0}\right) ^{2/3},
	\label{eq:app:thermodynamics}
\end{align}
where $c_0^2 \doteq  \gamma p_0/\rho_0$ and $u_0$ is the reference internal energy of the gas.  Hence, the invariants \eq{eq:app:J} are written as
\begin{equation}
	J_\pm = u \pm 3 c = \const.
\end{equation}
To determine the constant, let us assume a rarefraction wave propagating towards a gas whose density, pressure, and sound speed $(\rho_0,p_0,c_0)$ initially occupy the half space $x>0$.  At the edge of the propagating rarefraction wave, the fluid has not begun to move so $u=0$.  Hence,
\begin{equation}
	J_- = u - 3c = - 3 c_0.
\end{equation}
(The negative sign is chosen so that $u<0$, which means that the rarefraction wave expands towards the left.)  We can then write the local sound speed $c$ as a function of the local fluid velocity $u$:
\begin{equation}
	c = c_0 - u/3 .
	\label{eq:app:c}
\end{equation}
Using the thermodynamic relations \eq{eq:app:thermodynamics}, we can also write the fluid density and pressure as functions of $u$:
\begin{gather}
	\rho = \rho_0 \left( 1 +\frac{u}{3c_0} \right)^{3}, 
	\qquad
	p = p_0 \left( 1+ \frac{u}{3c_0} \right)^{5}.
	\label{eq:app:rho_p}
\end{gather}
The fluid expands towards the vacuum region $(x<0)$.  At the left boundary of the rarefraction wave, the fluid density and pressure are zero.  Hence, the edge of the rarefraction wave will travel at three times the speed of sound of the unperturbed fluid
\begin{equation}
	|u|_{\rm max} = 3 c_0.
\end{equation}

When combining \Eqs{eq:app:u_aux} and \eq{eq:app:c}, we obtain
\begin{equation}
	u 	=	 \frac{3}{4} \left( \xi - c_0  \right)
		=	 \frac{3}{4} c_0 \left( \frac{x}{c_0 t}	- 1 \right).
	\label{eq:app:u}
\end{equation}
The left boundary of the rarefraction wave facing the vacuum travels at a velocity $u=-3c_0$, so from the equation above, the left front of the rarefraction wave is given by $x=-3c_0t$.  At the right boundary of the rarefraction wave facing the fluid, one has $u=0$.  Therefore, the right front of the rarefraction wave is given by $x=c_0 t$.

Let us now calculate the mass fuel losses through the axial openings of a MagLIF liner.  The fuel mass lost $m_L(t)$ is defined as the total fluid mass that has passed the $x=0$ plane at a given time $t$.  This leads to
\begin{align}
	m_L(t) & \doteq 2\pi R_c^2 \int_{-3c_0 t}^0 \rho(t,x) \, \mathrm{d}x \notag \\
			&	=  2\pi R_c^2  \rho_0
					\int_{-3c_0 t}^0 \left[ 1+ \frac{1}{4}\left( \frac{x}{c_0 t}	- 1 \right)\right]^3
					 \, \mathrm{d}x \notag \\
			&	= 2\left( \frac{3}{4}\right)^4  \rho_0 (\pi R_c^2) t c_0 ,	\notag \\
			&	\simeq 0.63  \rho_0 (\pi R_c^2) t c_0 .				 
\end{align}
In the equation above, the factor $2$ accounts for the fact that the fuel can escape through the top and bottom ends of the liner. In the second line, we substituted \Eqs{eq:app:rho_p} and \eq{eq:app:u}.  We also introduced the transverse area $\pi R_c^2$ through which the fluid escapes.  We assume that the change of the transverse area and of the background fluid density and sound speed are slow compared to the propagation of the rarefraction wave into the fuel.  Thus, the resulting equation for the total fuel mass within the fuel cavity of a MagLIF liner is given by
\begin{equation}
	\frac{\mathrm{d} m }{\mathrm{d}t} = - 0.63 \pi \rho_0(t) c_0(t) R_c^2(t),
	\label{eq:app:m}
\end{equation}
Substituting $m=\pi R_{\rm in}^2 h \rho$ into \Eq{eq:app:m} leads to \Eq{eq:mass:rho}.

To estimate the corresponding energy losses due to the fuel outflow, we define $U_L(t)$ as the difference between the initial internal energy and the internal energy at time $t$ contained in the region $0\leq x \leq c_0t$, \ie where the rarefraction wave has perturbed the background fluid at rest.  For an ideal gas with $\gamma=5/3$, the internal energy per-unit-volume is $(3/2)p$.  Hence,
\begin{equation}
	U_L(t) \doteq 2 \left(\frac{3}{2}\right) \pi R_c^2 \int_0^{c_0 t} [p_0-p(t,x)] \, \mathrm{d}x \notag \\
\end{equation}
Upon substituting \Eq{eq:app:rho_p} and integrating, we find
\begin{align}
	U_L(t)  &	=  \left(\frac{3}{2}\right)  \left[ \frac{2}{3} + \left( \frac{3}{4}\right)^5 \right] p_0 (\pi R_c^2) t c_0 ,		
					\notag \\
			&	\simeq   0.90 \, \widehat{U}_0 \frac{R_c^2}{R_{\rm in}^2} t c_0 ,	
	\label{eq:app:internal_energy}			 
\end{align}
where we introduced the internal energy per-unit-length in the fuel such that $\smash{\widehat{U}_0} \doteq (3/2)p_0(\pi R_{\rm in}^2)$.  Taking the leading-order time derivative leads to \Eq{eq:basic:P_end}.


\begin{thebibliography}{83}
\bibitem[{foo({\natexlab{c}})}]{foot:mass_partition}
\bibinfo{note}{A different partitioning of the liner mass per-unit-length
  between the two regions is possible.}
\bibitem[{foo({\natexlab{d}})}]{foot:temp}
\bibinfo{note}{In \Eq{eq:liner:pfuel}, we have assumed that electrons and ions
  can be described using the same temperature $T$. This assumption is valid as
  long as the electron--ion energy exchange time is small compared to the
  characterstic times for the various phases of a MagLIF implosion; such as,
  preheat, implosion, and stagnation.}
\bibitem[{foo({\natexlab{e}})}]{foot:preheat}
\bibinfo{note}{In present-day MagLIF experiments, the laser pulse is $\sim
  6$-ns long compared to the $\sim100$-ns duration of the implosion. Due to the
  difference in timescales, it is a good approximation to consider that the
  preheat energy is delivered instantaneously. However, this approximation is
  not commonly assumed in more complex, radiation-magnetohydrodynamic models of
  MagLIF implosions.}
\bibitem[{foo({\natexlab{f}})}]{foot:alpha}
\bibinfo{note}{Equation \eq{eq:basic:P_alpha} assumes that the trapped $\alpha$
  particles instantaneously deposit their enegy to the electrons and that the
  subsequently heated electrons equilibrate instantaneously with the fuel ions.
  This is a good approximation as long as the characteristic time for these two
  processes ($\alpha$-particle slow down and electron--ion equilibration) are
  small compared to the characteristic fuel-burn time.}
\bibitem[{foo({\natexlab{h}})}]{foot:time_evaluation}
\bibinfo{note}{Note that, when evaluating the dimensionless quantities for the
  energy gain/loss mechanisms in
  \Eqs{eq:energy:Upsilon_alpha}--\eq{eq:energy:Upsilon_end}, we have used the
  characteristic plasma conditions for the preheat and stagnation phases given
  in \Tab{tab:conditions}. In particular, we have used the characteristic
  timescale $\Delta t$ of the implosion stage considered, not the
  characteristic timescale $t_\varphi$ of the voltage source.}
\bibitem[{foo({\natexlab{i}})}]{foot:nonessential}
\bibinfo{note}{It is commonly thought that small dimensionless parameters (much
  smaller than 0.1) describe physical effects that can be neglected. Therefore,
  such parameters are considered non-essential. It is worth noting here that
  large dimensionless parameters (much greater than 10) can also be considered
  non-essential since these parameters could be replaced by their reciprocals,
  which are also dimensionless and would be small.}
\end{thebibliography}
\end{document}